\documentclass[english]{IEEEtran}
\usepackage[T1]{fontenc}
\usepackage[latin9]{inputenc}
\usepackage{color}
\usepackage{babel}
\usepackage{array}
\usepackage{float}
\usepackage{multirow}
\usepackage{amsmath}
\usepackage{amsthm}
\usepackage{amssymb}
\usepackage{graphicx}
\usepackage{setspace}
\usepackage[unicode=true]
 {hyperref}

\makeatletter

\providecommand{\tabularnewline}{\\}
\floatstyle{ruled}
\newfloat{algorithm}{tbp}{loa}
\providecommand{\algorithmname}{Algorithm}
\floatname{algorithm}{\protect\algorithmname}

\theoremstyle{plain}
\newtheorem{thm}{\protect\theoremname}
\theoremstyle{plain}
\newtheorem{lem}[thm]{\protect\lemmaname}
\theoremstyle{plain}
\newtheorem{prop}[thm]{\protect\propositionname}

\usepackage{caption}

\newtheorem{assumption}{Assumption}

\usepackage{algpseudocode}
\algrenewcommand\textproc{} 

\usepackage{mathtools}

\newcommand\mystrut{\rule{0pt}{6.5pt}}

\usepackage{sepfootnotes}

\usepackage{etoolbox}

\newcommand{\algstrut}[1][\algruledefaultfactor]{\vrule width 0pt
depth .25\baselineskip height #1\baselineskip\relax}
\newcommand*{\algrule}[1][\algorithmicindent]{\hspace*{.5em}\vrule\algstrut
\hspace*{\dimexpr#1-.5em}}

\makeatletter
\newcount\ALG@printindent@tempcnta
\def\ALG@printindent{%
    \ifnum \theALG@nested>0
    \ifx\ALG@text\ALG@x@notext
    \else
    \unskip
    \ALG@printindent@tempcnta=1
    \loop
    \algrule[\csname ALG@ind@\the\ALG@printindent@tempcnta\endcsname]%
    \advance \ALG@printindent@tempcnta 1
    \ifnum \ALG@printindent@tempcnta<\numexpr\theALG@nested+1\relax
    \repeat
    \fi
    \fi
}%

\patchcmd{\ALG@doentity}{\noindent\hskip\ALG@tlm}{\ALG@printindent}{}{\errmessage{failed to patch}}

\AtBeginEnvironment{algorithmic}{\lineskip0pt}

\@ifundefined{showcaptionsetup}{}{%
 \PassOptionsToPackage{caption=false}{subfig}}
\usepackage{subfig}
\makeatother

\providecommand{\lemmaname}{Lemma}
\providecommand{\propositionname}{Proposition}
\providecommand{\theoremname}{Theorem}

\begin{document}
\title{A Fast Successive QP Algorithm for General Mean-Variance Portfolio
Optimization}
\author{Shengjie Xiu, Xiwen Wang, and Daniel P. Palomar, \IEEEmembership{Fellow, IEEE}\thanks{This work was supported by the Hong Kong GRF 16208917 and 16207019
research grants.}\thanks{The authors are with the Hong Kong University of Science and Technology
(HKUST), Clear Water Bay, Kowloon, Hong Kong (e-mail: \protect\href{mailto:sxiu@connect.ust.hk}{sxiu@connect.ust.hk};
\protect\href{mailto:xwangew@connect.ust.hk}{xwangew@connect.ust.hk};
\protect\href{mailto:palomar@ust.hk}{palomar@ust.hk}).}}
\maketitle
\begin{abstract}
The mean and variance of portfolio returns are the standard quantities
to measure the expected return and risk of a portfolio. Efficient
portfolios that provide optimal trade-offs between mean and variance
warrant consideration. To express a preference among these efficient
portfolios, investors have put forward many mean-variance portfolio
(MVP) formulations which date back to the classical Markowitz portfolio.
However, most existing algorithms are highly specialized to particular
formulations and cannot be generalized for broader applications. Therefore,
a fast and unified algorithm would be extremely beneficial. In this
paper, we first introduce a general MVP problem formulation that can
fit most existing cases by exploring their commonalities. Then, we
propose a widely applicable and provably convergent successive quadratic
programming algorithm (SCQP) for the general formulation. The proposed
algorithm can be implemented based on only the QP solvers and thus
is computationally efficient. In addition, a fast implementation is
considered to accelerate the algorithm. The numerical results show
that our proposed algorithm significantly outperforms the state-of-the-art
ones in terms of convergence speed and scalability.
\end{abstract}


\section{\textcolor{black}{Introduction\label{sec:Introduction} }}

Mean-variance analysis, pioneered by Harry Markowitz's publication
in 1952 \cite{markowitz1952portfolio}, is a breakthrough in modern
portfolio theory. It starts a new era of financial research using
quantitative tools and becomes one of the most widely-used investment
decision rules among academics and practitioners. Mean-variance analysis
assumes that the expected return and risk of a portfolio can be fully
measured by the mean and variance of the portfolio return. It accords
with the more general expected utility maximization when returns are
assumed to be normally distributed \cite{levy2004prospect}. In addition,
it is the theoretical support of the capital asset pricing model (CAPM)
developed by Sharpe \cite{sharpe1964capital} and Lintner \cite{lintner1965security}.

According to mean-variance analysis, rational risk-averse investors
would pursue efficient portfolios that provide the highest expected
return subject to a certain level of risk. The set of efficient portfolios
with different risk aversion levels is known as the efficient frontier.
However, the number of efficient portfolios is usually infinite, which
is of low applicability in practice. Therefore, to achieve a specific
efficient portfolio corresponding to individual risk appetite, investors
solve optimization problems based on well-designed trade-offs between
means and variances. These problems are collectively referred to as
mean-variance portfolio (MVP) problems, and their optimization process
is called MVP optimization \cite{fabozzi2007robust}.

Many MVP problem formulations have been proposed in the literature.
Markowitz portfolio, which optimizes a risk-adjusted return in the
form of a quadratic utility, serves as a starting point \cite{markowitz1952portfolio,markowitz1959portfolio}.
It has been popularly employed because it is simple for theoretical
analysis and numerical optimization. Nonetheless, this formulation
cannot fully characterize the preferences of a wide range of investors,
so it has been extended in several directions. First, the risk-adjusted
return is measured by other metrics like the Sharpe ratio \cite{sharpe1966ratio},
and the corresponding maximum Sharpe ratio portfolio (MSRP) also belongs
to the efficient frontier \cite{merton1972analytic}. Second, investors
adopt other utility functions, including exponential and logarithmic
ones \cite{kallberg1983utility}, to represent preferences over portfolio
returns. In this case, efficient portfolios could be obtained by applying
mean-variance approximations to expected utility \cite{levy1979mvApprox,levy2011capital}.
Third, practical linear and quadratic constraints, such as the restrictions
on desired return or risk, are included to take more investment guidelines
into account \cite{fabozzi2007robust,kolm2014review}. Since various
MVP formulations contain different problem structures, a broadly applicable
and numerically efficient algorithm is favored by academic research
and the financial industry. However, there are two significant challenges
in algorithm design.

\emph{Challenge \#1: Existing efficient algorithms are problem-dependent
and thus difficult to generalize.} Markowitz portfolio gains popularity
partly because it can be directly solved by efficient quadratic program
(QP) solvers. Nevertheless, involving more intricate functional forms
and constraints usually results in more complicated problems that
cannot benefit from the same computational advantage of QPs. For example,
given an MSRP which is a fractional program (FP) instead, we have
to resort to FP algorithms such as the bisection method, Dinkelbach's
algorithm \cite{dinkelbach1967}, and the quadratic transform \cite{shen2018fractional}.
However, these algorithms are limited to certain types of FPs and
cannot be applied to other problems \cite{zappone2015energy}. As
another example, given a worst-case robust global maximum return portfolio
(worst-case robust GMRP), the majorization-minimization (MM) algorithm
efficiently solves it via a sequence of quadratic upper-bound problems
\cite{sun2016majorization,scutari2018parallel}. Nonetheless, this
construction of quadratic bounds relies on the structure of $\ell_{2}$-norm,
and upper-bound problems may not be easily identifiable unless specific
convexity/concavity structures are contained in objective functions
\cite{razaviyayn2013unified,yang2016parallel}. Therefore, the scope
of applicability of MM is also limited.

\emph{Challenge \#2:} \emph{General algorithms are inefficient in
practice.} We could turn to traditional off-the-shelf optimization
methods to solve MVP problems. For example, feasible interior-point
methods (FIP) \cite{byrd1999interior} and feasible sequential quadratic
programming (FSQP) \cite{lawrence2001FSQP} are commonly used. However,
they are computationally expensive in most cases \cite{scutari2016parallel},
especially for nonconvex problems \cite{feng2015scrip}. Metaheuristic
methods can be applied, including differential evolution and genetic
algorithms. Nonetheless, they are inefficient as they require many
objective function evaluations and barely exploit problem-specific
information \cite{sivanandam2008genetic}. Although general solvers
are available for some MVPs, it is widely accepted that QP algorithms
are faster, more reliable, and easier to use.

In response to the challenges in designing an efficient and unified
algorithm for MVP optimization, this work presents the following major
contributions:
\begin{itemize}
\item We provide a general formulation that can characterize most of the
existing MVP formulations.
\item We propose an algorithm that solves different MVPs with various structures
by uniformly solving a sequence of QPs. It can reuse the well-developed
QP solvers, and has provable convergence and a fast implementation.
\item We present extensive results to show that our proposed algorithm is
more efficient and scalable than the state-of-the-art methods.
\end{itemize}
\begin{onehalfspace}
\begin{table*}[!ht]\caption{Formulations of well-researched MVPs under the framework of \ref{eq:General MVP formulation}.\protect \label{tab:Example portfolios}}\centering{}{\large{}}%
\begin{tabular}{|l|c|c|c|}
\hline 
\multicolumn{1}{|c|}{\textbf{Portfolio}} & \textbf{$F(\mathbf{x}(\mathbf{w}),\mathbf{y}(\mathbf{w}))$} & $\mathbf{g}(\mathbf{w})\le\mathbf{0}$ & \textbf{Problem class}\tabularnewline
\hline 
Markowitz portfolio & $-x(\mathbf{w})+\frac{\alpha}{2}y(\mathbf{w})$ & - & QP\tabularnewline
\hline 
\multirow{2}{*}{MSRP} & $-(x(\mathbf{w})-r_{f})/\sqrt{y(\mathbf{w})}$ & - & \multirow{3}{*}{FP}\tabularnewline
\cline{2-3} \cline{3-3} 
 & $\sqrt{y(\mathbf{w})}/(x(\mathbf{w})-r_{f})$ & - & \tabularnewline
\cline{1-3} \cline{2-3} \cline{3-3} 
MGSRP & $-(x(\mathbf{w})-r_{f})/y(\mathbf{w})^{\beta}$ & - & \tabularnewline
\hline 
Worst-case robust GMRP & $-x(\mathbf{w})+\alpha\sqrt{y(\mathbf{w})}$ & - & SOCP\tabularnewline
\hline 
Expected utility portfolio & $-U(x(\mathbf{w}))-\frac{1}{2}U^{''}(x(\mathbf{w}))y(\mathbf{w})$ & - & Depends on $U$\tabularnewline
\hline 
Kelly portfolio & $-\log(1+x(\mathbf{w}))+\frac{1}{2}\frac{y(\mathbf{w})}{(1+x(\mathbf{w}))^{2}}$ & - & {\large{}-}\tabularnewline
\hline 
Return-constrained Markowitz portfolio & $y(\mathbf{w})$ & $x(\mathbf{w})\ge a$ & QP\tabularnewline
\hline 
\multirow{1}{*}{Risk-constrained Markowitz portfolio} & $-x(\mathbf{w})$ & $y(\mathbf{w})\le b$ & \multirow{1}{*}{QCQP}\tabularnewline
\hline 
\end{tabular}{\large\par}

\end{table*}
\end{onehalfspace}

The rest of the paper is organized as follows. We begin with the general
MVP formulation and its motivating examples in Section \ref{sec:formulation}.
Then, in Section \ref{sec:Proposed-Algorithm}, we propose our successive
QP algorithm for the general problem formulation. The detailed analysis
of the proposed algorithm is provided in Section \ref{sec:Analysis},
and the fast implementation that exploits the underlying sparsity
pattern is developed in Section \ref{sec:Fast}. Section \ref{sec:Applications-and-Experiments}
justifies the proposed algorithm's broad applicability and performance
with comprehensive experiments. Finally, Section \ref{sec:conclusion}
concludes this paper.

\section{Problem Formulation\label{sec:formulation}}

Firstly, we introduce the notation and related background. We denote
by $\mathbf{r}\in\mathbb{R}^{N}$ the returns of $N$ assets. The
mean vector and covariance matrix of the returns are represented by
$\boldsymbol{\mu}\in\mathbb{R}^{N}$ and $\boldsymbol{\Sigma}\in\mathbb{S}_{++}^{N}$,
respectively. We let $\mathbf{w}\in\mathbb{R}^{N}$ denote the portfolio
weights. Then, the expected return and risk of this portfolio are
individually measured by the mean ($\mathbf{w}^{\intercal}\boldsymbol{\mu}$)
and variance ($\mathbf{w}^{\intercal}\boldsymbol{\Sigma}\mathbf{w}$)
of the portfolio return ($\mathbf{w}^{\intercal}\mathbf{r}$). 

The classical Markowitz portfolio can be obtained by solving the following
QP problem:
\begin{equation}
\underset{\mathbf{w}\in\mathcal{W}}{\mathsf{minimize}}\;\;-\mathbf{w}^{\intercal}\boldsymbol{\mu}+\frac{\alpha}{2}\mathbf{w}^{\intercal}\boldsymbol{\Sigma}\mathbf{w},\label{eq:standard MVP}
\end{equation}
where $\alpha\ge0$ is a risk aversion parameter that determines the
trade-off between expected return and risk. $\mathcal{W}$ is the
convex feasible set of $\mathbf{w}$ and is defined in its simplest
form as
\begin{equation}
\mathcal{W}=\left\{ \mathbf{w}\ge\mathbf{0},\:\mathbf{1}^{\intercal}\mathbf{w}=1\right\} ,\label{eq:W definition}
\end{equation}
which includes the long-only and capital budget constraints \cite{markowitz1959portfolio}.
In practice, this set may contain other constraints such as turnover
control, upper bound limits, leverage constraints, etc. For illustrative
purposes, we focus on the constraints in (\ref{eq:W definition}).

The original formulation proposed by Markowitz has been extended in
two aspects. The first one is amending the mean-variance framework
with a broad class of objective functions based on different combinations
of means and variances to express investors\textquoteright{} preferences.
In practice, we may have multiple estimates of $\boldsymbol{\mu}$
and $\boldsymbol{\Sigma}$, denoted as $\{\boldsymbol{\mu}_{1},\boldsymbol{\mu}_{2},\cdots,\boldsymbol{\mu}_{p}\}$
and $\{\boldsymbol{\Sigma}_{1},\boldsymbol{\Sigma}_{2},\cdots,\boldsymbol{\Sigma}_{q}\}$,
by using different estimation methods or market regimes. Therefore,
we can obtain multiple estimates of expected return and risk, denoted
as vector-valued functions $\mathbf{x}(\mathbf{w}):\mathcal{W}\rightarrow\mathbb{R}^{p}$
and $\mathbf{y}(\mathbf{w}):\mathcal{W}\rightarrow\mathbb{R}_{++}^{q}$,
\begin{equation}
\begin{aligned}\mathbf{x}\left(\mathbf{w}\right)= & \left[x_{1}\left(\mathbf{w}\right),\dots,x_{p}\left(\mathbf{w}\right)\right],\\
\mathbf{y}\left(\mathbf{w}\right)= & \left[y_{1}\left(\mathbf{w}\right),\dots,y_{q}\left(\mathbf{w}\right)\right],
\end{aligned}
\label{eq:mean and variance terms}
\end{equation}
where 
\begin{equation}
x_{i}\left(\mathbf{w}\right)=\mathbf{w}^{\intercal}\boldsymbol{\mu}_{i},\quad y_{j}\left(\mathbf{w}\right)=\mathbf{w}^{\intercal}\boldsymbol{\Sigma}_{j}\mathbf{w}.
\end{equation}
To represent the family of objective functions of means $\mathbf{x}(\mathbf{w})$
and variances $\mathbf{y}(\mathbf{w})$, we consider such functions
in their most general form $F(\mathbf{x},\mathbf{y}):\mathbb{R}^{p}\times\mathbb{R}^{q}\rightarrow\mathbb{R}$.
The second aspect consists in adding minimum expected return or bearable
maximum risk as practical constraints.

Therefore, the general MVP can be formulated as\begin{equation}
\begin{aligned}
&\,\,\underset{\mathbf{w}}{\mathsf{minimize}}\,\,\, \quad\quad f\left(\mathbf{w}\right)\triangleq F\left(\mathbf{x}\left(\mathbf{w}\right),\mathbf{y}\left(\mathbf{w}\right)\right)\\
&\begin{rcases}
\mathsf{subject\,\,to} & \quad x_{i}\left(\mathbf{w}\right)\ge a_{i},\,\,i=1,\dots,p\\  
& \quad y_{j}\left(\mathbf{w}\right)\le b_{j},\,\,j=1,\dots,q\\  
& \quad\mathbf{w}\in\mathcal{W}, 
\end{rcases}\triangleq\mathcal{K}
\end{aligned}
\tag{$\mathcal{P}$}
\label{eq:General MVP formulation}
\end{equation}where
\begin{itemize}
\item $f\left(\mathbf{w}\right):\mathcal{K}\rightarrow\mathbb{R}$ is a
continuously differentiable and possibly nonconvex function.
\item $\mathcal{K}$ is the feasible set that contains at least one strictly
feasible point, where $a_{i}$ is the lower limit on the expected
return, and $b_{j}$ is the upper limit on the risk. The constraints
on the expected return and risk are named mean-variance constraints.
They can be written in a compact vector-valued function $\mathbf{g}(\mathbf{w}):\mathcal{W}\rightarrow\mathbb{R}^{p}\times\mathbb{R}^{q}$
satisfying
\begin{equation}
\mathbf{g}\left(\mathbf{w}\right)\triangleq\left[\begin{array}{c}
\mathbf{g}_{x}\left(\mathbf{w}\right)\\
\mathbf{g}_{y}\left(\mathbf{w}\right)
\end{array}\right]\le\mathbf{0},\label{eq:compact g constraint}
\end{equation}
with
\begin{equation}
\left[\mathbf{g}_{x}\left(\mathbf{w}\right)\right]_{i}\triangleq a_{i}-x_{i}\left(\mathbf{w}\right),\:\left[\mathbf{g}_{y}\left(\mathbf{w}\right)\right]_{j}\triangleq y_{j}\left(\mathbf{w}\right)-b_{j}.
\end{equation}
\end{itemize}
Compared with Problem (\ref{eq:standard MVP}), which well-developed
QP solvers can efficiently solve, \ref{eq:General MVP formulation}
is usually more complicated because of the substantial flexibility
regarding the objective function $f$ (i.e., $F$) and the inclusion
of mean-variance constraints. Therefore, our goal is to efficiently
deal with \ref{eq:General MVP formulation} by solving Problem (\ref{eq:standard MVP})
iteratively. 

As the risk-return trade-off characterized by $F$ should be reasonable,
we require the following natural assumption.

\begin{assumption} For each $x_{i}$ and $y_{j}$ that exists in
$F$, we have
\[
\nabla_{x_{i}}F\left(\mathbf{x},\mathbf{y}\right)<0,\quad\nabla_{y_{j}}F\left(\mathbf{x},\mathbf{y}\right)>0.
\]
where $\nabla_{x_{i}}F(\mathbf{x},\mathbf{y})$ and $\nabla_{y_{j}}F(\mathbf{x},\mathbf{y})$
denote the partial gradient of $F$ evaluated at $x_{i}$ and $y_{j}$,
respectively.\footnotemark[1]\footnotetext[1]{$x_{i}$ (or $y_{j}$) exists in $F$ iff $\nabla_{x_{i}}F$ (or $\nabla_{y_{j}}F$) does not always equal to $0$.} 

\label{ASMP: Pareto}\end{assumption}

Assumption \ref{ASMP: Pareto} implies that function $F(\mathbf{x},\mathbf{y})$
is decreasing with respect to $x_{i}$ and increasing with respect
to $y_{j}$. In other words, a higher expected return is always better,
while more risk is always worse. It is a direct generalization of
the continuity--monotonicity--finiteness axiom of the two-moment
decision model \cite{johnstone2013mean}.

\subsection{Motivating Examples of General MVP\label{subsec:motivating examples}}

The general MVP formulation \ref{eq:General MVP formulation} can
be easily customized to many well-researched portfolios. Representative
portfolios and their classes of optimization problems are listed in
Table \ref{tab:Example portfolios} and elaborated next.\footnotemark[2]\footnotetext[2]{Scalar functions $x(\mathbf{w})$ and $y(\mathbf{w})$ denote a single expected return and a single risk of a portfolio, respectively.} 

\subsubsection{On objective function $F$}

Investors' preferences for different risk-return trade-off strategies
can be modeled with different $F$. We provide some examples as follows.

\textbf{MSRP}: Sharpe ratio evaluates the expected return earned over
the risk-free rate $r_{f}$ per unit of volatility \cite{sharpe1966ratio}.
To find the portfolio that provides the maximum Sharpe ratio, we minimize
the negative Sharpe ratio written as $F_{\text{SR}}$,
\begin{equation}
F_{\text{SR}}\left(x\left(\mathbf{w}\right),y\left(\mathbf{w}\right)\right)=-\frac{x\left(\mathbf{w}\right)-r_{f}}{\sqrt{y\left(\mathbf{w}\right)}},\label{eq:msrp form}
\end{equation}
or, alternatively, the inverse of Sharpe ratio. MSRP is a special
case of the maximum generalized Sharpe ratio portfolio (MGSRP) introduced
in \cite{landsman2018generalized} with the objective function $F_{\text{GSR}}$,
\begin{equation}
F_{\text{GSR}}\left(x\left(\mathbf{w}\right),y\left(\mathbf{w}\right)\right)=-\frac{x\left(\mathbf{w}\right)-r_{f}}{y\left(\mathbf{w}\right)^{\beta}},\label{eq:mgsr form}
\end{equation}
where $\beta\ge1/2$ is a risk aversion parameter. The corresponding
optimization problems are all FPs.

\textbf{Worst-case robust GMRP}: The estimate of the expected return
$\hat{\boldsymbol{\mu}}$ is inevitably subject to estimation error.
The true $\boldsymbol{\mu}$ may be assumed to fall within an uncertainty
ellipsoid shaped by $\boldsymbol{\Sigma}$, i.e., $\mathcal{U}_{\boldsymbol{\mu}}=\{\boldsymbol{\mu}=\hat{\boldsymbol{\mu}}+\alpha\boldsymbol{\Sigma}^{1/2}\mathbf{u}|\|\mathbf{u}\|_{2}\le1\}$,
where $\alpha>0$ is a predefined parameter. Then, the objective function
of the worst-case robust GMRP is written as $f_{\text{WC}}\left(\mathbf{w}\right)=-\mathsf{min}_{\boldsymbol{\mu}\in\mathcal{U}_{\boldsymbol{\mu}}}\,\mathbf{w}^{\intercal}\boldsymbol{\mu}$,
which can be reformulated as the following $F_{\text{WC}}$ according
to \cite{lobo2000worstcase},
\begin{equation}
F_{\text{WC}}\left(x\left(\mathbf{w}\right),y\left(\mathbf{w}\right)\right)=-x\left(\mathbf{w}\right)+\alpha\sqrt{y\left(\mathbf{w}\right)}.\label{eq:worst-case GMRP}
\end{equation}
The resulting optimizing problem can be recast as a second-order cone
program (SOCP).

\textbf{Expected utility portfolio}: A utility function $U$ measures
the relative satisfaction with portfolio return. In this setting,
the expected utility portfolio is attained by maximizing the expected
utility, i.e.,
\begin{equation}
\underset{\mathbf{w}\in\mathcal{W}}{\mathsf{maximize}}\;\;\mathbb{E}\left[U\left(\mathbf{w}^{\intercal}\mathbf{r}\right)\right].
\end{equation}
It is common and effective to apply mean-variance approximations to
the expected utility \cite{levy1979mvApprox,markowitz2014mean}, i.e.,
performing Taylor expansion at the point $\mathbf{r}=\boldsymbol{\mu}$
and ignoring the moments greater than the second one:
\begin{equation}
\begin{aligned}\mathbb{E}\left[U\left(\mathbf{w}^{\intercal}\mathbf{r}\right)\right] & \approx U\left(\mathbf{w}^{\intercal}\boldsymbol{\mu}\right)+U^{'}\left(\mathbf{w}^{\intercal}\boldsymbol{\mu}\right)\mathbb{E}\left[\mathbf{w}^{\intercal}\left(\mathbf{r}-\boldsymbol{\mu}\right)\right]\\
 & \quad+\frac{1}{2}U^{''}\left(\mathbf{w}^{\intercal}\boldsymbol{\mu}\right)\mathbb{E}\left[\left(\mathbf{w}^{\intercal}\left(\mathbf{r}-\boldsymbol{\mu}\right)\right)^{2}\right]\\
 & =U\left(\mathbf{w}^{\intercal}\boldsymbol{\mu}\right)+\frac{1}{2}U^{''}\left(\mathbf{w}^{\intercal}\boldsymbol{\mu}\right)\mathbf{w}^{\intercal}\boldsymbol{\Sigma}\mathbf{w}.
\end{aligned}
\label{eq:EU MV-approx}
\end{equation}
Therefore, the approximated expected utility as the objective function
can be expressed as $F_{\text{EU}}$,
\begin{align}
F_{\text{EU}} & \left(x\left(\mathbf{w}\right),y\left(\mathbf{w}\right)\right)\nonumber \\
 & =-U\left(x\left(\mathbf{w}\right)\right)-\frac{1}{2}U^{''}\left(x\left(\mathbf{w}\right)\right)y\left(\mathbf{w}\right).\label{eq:EU MV-approx f}
\end{align}

\textbf{Kelly  portfolio}: To achieve maximum growth of wealth, Kelly
portfolio maximizes the expected value of the logarithm of portfolio
return \cite{kelly1956new,thorp1975portfolio}. It coincides with
the expected utility portfolio using the logarithmic utility function
$U(\mathbf{w}^{\intercal}\boldsymbol{r})=\log(1+\mathbf{w}^{\intercal}\boldsymbol{r})$.
Accordingly, the objective function $F_{\text{KL}}$ of the Kelly
portfolio is given by
\begin{align}
F_{\text{KL}} & \left(x\left(\mathbf{w}\right),y\left(\mathbf{w}\right)\right)\nonumber \\
 & =-\log\left(1+x\left(\mathbf{w}\right)\right)+\frac{1}{2}\frac{y\left(\mathbf{w}\right)}{\left(1+x\left(\mathbf{w}\right)\right)^{2}},
\end{align}
following the approximation (\ref{eq:EU MV-approx f}), as shown in
\cite{pulley1983mean}.

\subsubsection{On mean-variance constraints}

The mean-variance constraints present in the following examples.

\textbf{Alternatives of Markowitz portfolio}: Markowitz portfolio
(\ref{eq:standard MVP}) has two alternative formulations, which lead
to the same efficient frontier. One minimizes the risk given a lower
limit on the return, known as a return-constrained Markowitz portfolio:
\begin{equation}
\begin{aligned}\underset{\mathbf{w}\in\mathcal{W}}{\mathsf{minimize}}\,\,\, & \quad y\left(\mathbf{w}\right)\\
\mathsf{subject\,\,to} & \quad x\left(\mathbf{w}\right)\ge a.
\end{aligned}
\label{eq:return-const mark}
\end{equation}
The other one maximizes the return given an upper limit on the risk,
known as a risk-constrained Markowitz portfolio. It is formulated
as the quadratically constrained quadratic program (QCQP):
\begin{equation}
\begin{aligned}\underset{\mathbf{w}\in\mathcal{W}}{\mathsf{maximize}}\,\, & \quad x\left(\mathbf{w}\right)\\
\mathsf{subject\,\,to} & \quad y\left(\mathbf{w}\right)\le b.
\end{aligned}
\label{eq:risk-const mark}
\end{equation}

We have seen that the general formulation \ref{eq:General MVP formulation}
has a number of instances whose problem structures are distinct from
each other. However, their relationship with a common QP representation
can be explored, thus leading to an efficient successive QP algorithm
introduced in the next section.

\section{Proposed Algorithm\label{sec:Proposed-Algorithm}}

In this section, we propose an algorithm to solve \ref{eq:General MVP formulation}
via a sequence of QP surrogate problems.

\subsection{Algorithmic Framework}

\begin{onehalfspace}
\begin{table*}[!ht]\caption{Closed-form updates of $\boldsymbol{\lambda}$ and $\boldsymbol{\eta}$ for well-researched MVPs.  \label{tab:Example algorithms}}\centering{}

{\large{}}%
\begin{tabular}{|l|l|l|}
\hline 
\multicolumn{1}{|c|}{\textbf{Portfolio}} & \multicolumn{2}{c|}{\textbf{Updates of} $\boldsymbol{\lambda}$ \textbf{and} $\boldsymbol{\eta}$}\tabularnewline
\hline 
Markowitz portfolio & $\lambda_{x}\gets1$ & $\lambda_{y}\gets\alpha/2$\tabularnewline
\hline 
MSRP\footnotemark[3] & $\lambda_{x}\gets1$ & $\lambda_{y}\gets(x^{k}-r_{f})/(2y^{k})$\tabularnewline
\hline 
GMSRP\footnotemark[3] & $\lambda_{x}\gets1$ & $\lambda_{y}\gets\beta(x^{k}-r_{f})/y^{k}$\tabularnewline
\hline 
Worst-case robust GMRP & $\lambda_{x}\gets1$ & $\lambda_{y}\gets\alpha/(2\sqrt{y^{k}})$\tabularnewline
\hline 
Expected utility portfolio & $\lambda_{x}\gets U^{'}(x^{k})+U^{'''}(x^{k})y^{k}/2$, & $\lambda_{y}\gets-U^{''}(x^{k})/2$\tabularnewline
\hline 
Kelly portfolio & $\lambda_{x}\gets1/(1+x^{k})+y^{k}/(1+x^{k})^{3}$ & $\lambda_{y}\gets1/(2(1+x^{k})^{2})$\tabularnewline
\hline 
Return-constrained Markowitz portfolio & $\eta_{x}\gets\left[\eta_{x}+\alpha\left(a-x\left(\mathbf{w}^{\star}\left(\lambda_{y},\eta_{x}\right)\right)\right)\right]_{+}$ & $\lambda_{y}\gets1$,\tabularnewline
\hline 
Risk-constrained Markowitz portfolio & $\eta_{y}\gets\left[\eta_{y}+\alpha\left(y\left(\mathbf{w}^{\star}\left(\lambda_{x},\eta_{y}\right)\right)-b\right)\right]_{+}$ & $\lambda_{x}\gets1$,\tabularnewline
\hline 
\end{tabular}{\large\par}

\end{table*}
\end{onehalfspace}

Unlike \ref{eq:General MVP formulation} which implicitly specifies
investors' risk-return trade-off, the QP surrogate problem that we
are interested in applies a straightforward characterization of this
trade-off. It adopts the following formulation 
\begin{equation}
\underset{\mathbf{w}\in\mathcal{W}}{\mathsf{minimize}}\;\;-\left(\boldsymbol{\lambda}_{x}+\boldsymbol{\eta}_{x}\right)^{\intercal}\mathbf{x}\left(\mathbf{w}\right)+\left(\boldsymbol{\lambda}_{y}+\boldsymbol{\eta}_{y}\right)^{\intercal}\mathbf{y}\left(\mathbf{w}\right),\label{eq:QP surrogate}
\end{equation}
whose objective function is a weighted combination of the means and
variances. Specifically, $\boldsymbol{\lambda}=[\boldsymbol{\lambda}_{x};\boldsymbol{\lambda}_{y}]$
is the non-negative weight that characterizes the mean-variance trade-off
reflected in $f$, and $\boldsymbol{\eta}=[\boldsymbol{\eta}_{x};\boldsymbol{\eta}_{y}]$
is the non-negative weight that controls the magnitude of impact from
$\mathbf{g}(\mathbf{w})\le\mathbf{0}$. Without loss of generality,
we assume that Problem (\ref{eq:QP surrogate}) has a unique solution
$\hat{\mathbf{w}}(\boldsymbol{\lambda},\boldsymbol{\eta})$ at every
iterate in our algorithm. By verifying the optimality conditions,
$\hat{\mathbf{w}}(\boldsymbol{\lambda},\boldsymbol{\eta})$ coincides
with the stationary solution of \ref{eq:General MVP formulation}
when $\boldsymbol{\lambda}$ and $\boldsymbol{\eta}$ are correctly
chosen (see Section \ref{sec:Analysis}). Therefore, instead of solving
the complicated \ref{eq:General MVP formulation} directly, we propose
that by solving a sequence of QP surrogate problems (\ref{eq:QP surrogate}),
the weights could be dynamically adjusted such that $\hat{\mathbf{w}}(\boldsymbol{\lambda},\boldsymbol{\eta})$
converges to the stationary solution of \ref{eq:General MVP formulation}. 

As mentioned, we already have efficient QP solvers for Problem (\ref{eq:standard MVP}).
Interestingly, Problem (\ref{eq:QP surrogate}) can be rewritten in
the form of (\ref{eq:standard MVP}) as follows

\begin{equation}
\underset{\mathbf{w}\in\mathcal{W}}{\mathsf{minimize}}\;\;-\mathbf{w}^{\intercal}\bar{\boldsymbol{\mu}}+\frac{1}{2}\mathbf{w}^{\intercal}\bar{\boldsymbol{\Sigma}}\mathbf{w},\label{eq:surrogate Mark form}
\end{equation}
where
\begin{equation}
\bar{\boldsymbol{\mu}}=\sum_{i=1}^{p}\left(\lambda_{x,i}+\eta_{x,i}\right)\boldsymbol{\mu}_{i},\enskip\bar{\boldsymbol{\Sigma}}=\sum_{j=1}^{q}2\left(\lambda_{y,j}+\eta_{y,j}\right)\boldsymbol{\Sigma}_{j}.
\end{equation}
Therefore, solving Problem (\ref{eq:QP surrogate}) shares the same
computational convenience as dealing with Problem (\ref{eq:standard MVP}).
Overall, this algorithmic framework can be efficient due to the low
computational cost of QP surrogate problems with the undermentioned
fast implementation presented in Section \ref{sec:Fast} and the simplicity
of the weight updates introduced next.

We begin with a systematic manner that updates $\boldsymbol{\lambda}$
by iteratively approximating $f$ using quadratic functions $\tilde{f}$.
Given the current iterate $\mathbf{w}^{k}$, according to the successive
convex approximation (SCA) framework \cite{scutari2016parallel,scutari2013decomposition},
we propose to optimize \ref{eq:General MVP formulation} by iteratively
solving the following problem
\begin{align}
\underset{\mathbf{w}\in\mathcal{K}}{\mathsf{minimize}}\;\,\, & \tilde{f}\left(\mathbf{w};\mathbf{w}^{k}\right)\triangleq\nabla_{\mathbf{x}}F\left(\mathbf{x}^{k},\mathbf{y}^{k}\right)^{\intercal}\mathbf{x}\left(\mathbf{w}\right)\nonumber \\
 & \qquad\qquad\qquad+\nabla_{\mathbf{y}}F\left(\mathbf{x}^{k},\mathbf{y}^{k}\right)^{\intercal}\mathbf{y}\left(\mathbf{w}\right),\label{eq:f app}
\end{align}
where we denote $\mathbf{x}^{k}=\mathbf{x}(\mathbf{w}^{k})$ and $\mathbf{y}^{k}=\mathbf{y}(\mathbf{w}^{k})$
for notational simplicity. If $\mathcal{K}=\mathcal{W}$, i.e., the
mean-variance constraints are not present, Problem (\ref{eq:f app})
coincides with Problem (\ref{eq:QP surrogate}) when $\boldsymbol{\lambda}$
is chosen as
\begin{equation}
\boldsymbol{\lambda}_{x}^{k}=-\nabla_{\mathbf{x}}F\left(\mathbf{x}^{k},\mathbf{y}^{k}\right),\:\:\boldsymbol{\lambda}_{y}^{k}=\nabla_{\mathbf{y}}F\left(\mathbf{x}^{k},\mathbf{y}^{k}\right),\label{eq:lambda update}
\end{equation}
and $\boldsymbol{\eta}$ is ignored. In this way, \ref{eq:General MVP formulation}
can be handled via a sequence of QP problems. However, if quadratic
constraints are included, Problem (\ref{eq:f app}) is still a QCQP,
which has a much higher complexity than a QP.
\begin{algorithm}
\caption{SCQP Algorithm for \ref{eq:General MVP formulation}. \label{alg:successive QP}}

\textbf{Input:} $k=0$, $\mathbf{w}^{0}\in\mathcal{K}$, $\boldsymbol{\eta}^{0}\ge\mathbf{0}$,
and $\{\alpha^{l}\},\{\gamma^{k}\}\in(0,1]$;

\begin{algorithmic}[1]

\Repeat

\State $\boldsymbol{\lambda}_{x}^{k}=-\nabla_{\mathbf{x}}F\left(\mathbf{x}^{k},\mathbf{y}^{k}\right)$,
$\boldsymbol{\lambda}_{y}^{k}=\nabla_{\mathbf{y}}F\left(\mathbf{x}^{k},\mathbf{y}^{k}\right)$;

\If{$\mathcal{K}=\mathcal{W}$ (no mean-variance constraints)}

\State Compute $\hat{\mathbf{w}}^{k}=\hat{\mathbf{w}}(\boldsymbol{\lambda}^{k},\mathbf{0})$
via QP (\ref{eq:QP surrogate});

\Else

\State Set $l=0;$.

\Repeat

\State Compute $\hat{\mathbf{w}}^{k}=\hat{\mathbf{w}}(\boldsymbol{\lambda}^{k},\boldsymbol{\eta}^{l})$
via QP (\ref{eq:QP surrogate});

\State $\boldsymbol{\eta}^{l+1}=[\boldsymbol{\eta}^{l}+\alpha^{l}\mathbf{g}(\hat{\mathbf{w}}^{k})]_{+}$;

\State $l\gets l+1$;

\Until convergence

\EndIf

\State $\mathbf{w}^{k+1}=\mathbf{w}^{k}+\gamma^{k}(\hat{\mathbf{w}}^{k}-\mathbf{w}^{k})$;

\State $k\gets k+1$;

\Until convergence

\end{algorithmic}

\textbf{Output:} A stationary solution $\mathbf{w}^{k+1}$ of \ref{eq:General MVP formulation}.
\end{algorithm}

 To further recast Problem (\ref{eq:f app}) as a sequence of QP
(\ref{eq:QP surrogate}), we apply the partial relaxation method \cite{palomar2007alternative}.
Given Problem (\ref{eq:f app}) with $\boldsymbol{\lambda}=\boldsymbol{\lambda}^{k}$,
we relax its mean-variance constraints and then optimize its Lagrangian
dual problem
\begin{equation}
\underset{\boldsymbol{\eta}\ge\mathbf{0}}{\mathsf{maximize}}\;\;h\left(\boldsymbol{\eta};\boldsymbol{\lambda}^{k}\right),\label{eq:partial dual problem}
\end{equation}
where $h$ is the Lagrangian dual function of (\ref{eq:f app}), and
$\boldsymbol{\eta}$ is the Lagrange multiplier with respect to $\mathbf{g}(\mathbf{w})\le\mathbf{0}$.
At the $l$th iteration, the gradient projection method updates $\boldsymbol{\eta}^{l+1}$
by
\begin{equation}
\boldsymbol{\eta}^{l+1}=\left[\boldsymbol{\eta}^{l}+\alpha^{l}\nabla h(\boldsymbol{\eta}^{l};\boldsymbol{\lambda}^{k})\right]_{+},\label{eq:eta update}
\end{equation}
where $\alpha^{l}$ is a step-size and $[\cdot]_{+}=\mathsf{max}\{0,\cdot\}$
denotes projection for nonnegative constraints. Following \cite{bertsekas1999nonlinear},
the gradient of $h$ can be computed as
\begin{equation}
\nabla h(\boldsymbol{\eta}^{l};\boldsymbol{\lambda}^{k})=\mathbf{g}(\hat{\mathbf{w}}(\boldsymbol{\lambda}^{k},\boldsymbol{\eta}^{l})).\label{eq:grad h with QP}
\end{equation}
Obviously, each evaluation of $\nabla h$ is attained by a QP (\ref{eq:QP surrogate}).
Therefore, Problem (\ref{eq:f app}) is solved by alternatively conducting
a QP (\ref{eq:QP surrogate}) to compute $\nabla h$ and taking an
inexpensive gradient projection step.

Algorithm \ref{alg:successive QP} summarizes the proposed algorithm
for \ref{eq:General MVP formulation} that consists of solving a sequence
of QP (\ref{eq:QP surrogate}), which is referred to as SCQP (SucCessive
QP). It adopts a double-loop scheme: the outer loop decomposes \ref{eq:General MVP formulation}
into problems (\ref{eq:f app}) with quadratic objectives and iteratively
updates $\boldsymbol{\lambda}$; the inner loop further recasts Problem
(\ref{eq:f app}) into QP surrogate problems (\ref{eq:QP surrogate})
and iteratively updates $\boldsymbol{\eta}$\@. We remark again that
Algorithm \ref{alg:successive QP} can be generalized when $\mathcal{W}$
includes other constraints. Technical details of SCQP are deferred
to Section \ref{sec:Analysis}.

Algorithm \ref{alg:successive QP} can be simplified in two scenarios.
First, if $\mathcal{K}=\mathcal{W}$, i.e., \ref{eq:General MVP formulation}
does not include mean-variance constraints, we get rid of the difficulty
mentioned in solving Problem (\ref{eq:f app}), and thus we can set
$\boldsymbol{\eta}$ to zero and ignore its update. Second, if the
objective function $f$ already satisfies the form of $\tilde{f}$,
we maintain $\boldsymbol{\lambda}$ without tuning. Though \ref{eq:General MVP formulation}
allows different problem structures that result in various difficulties
in numerical optimization, with the proposed algorithm, we only require
to focus on QP surrogate problems with simple updates of $\boldsymbol{\lambda}$
and $\boldsymbol{\eta}$. The applicability of the proposed algorithm
will be described in detail in the next subsection. 

\subsection{Applicability of Proposed Algorithm}

The proposed algorithm is broadly applicable in two aspects. The
first aspect is that when dealing with different instances of \ref{eq:General MVP formulation},
the modification only resides in the updates of $\boldsymbol{\lambda}$
and $\boldsymbol{\eta}$. Some closed-form updates are summarized
in Table \ref{tab:Example algorithms}, and we note that they are
easily derived and almost computationally free. The second aspect
is that SCQP, as a more general algorithmic framework, includes some
problem-dependent algorithms as specific cases.\footnotetext[3]{$\boldsymbol{\lambda}$ is scaled by a constant factor so that $\lambda_{x}\leftarrow1$.} 

\subsubsection*{On the connection to quadratic transform for MSRP}

Under Assumption \ref{ASMP: Pareto}, the MSRP problem is equivalent
to\footnotemark[4]\footnotetext[4]{The satisfaction of Assumption \ref{ASMP: Pareto} indicates $x(\mathbf{w})>r_{f}$.} 
\begin{equation}
\underset{\mathbf{w}\in\mathcal{W}}{\mathsf{minimize}}\;\;-\frac{\left(x\left(\mathbf{w}\right)-r_{f}\right)^{2}}{y\left(\mathbf{w}\right)}.\label{eq:equivalent msrp}
\end{equation}
Problem (\ref{eq:equivalent msrp}) can be solved by the quadratic
transform \cite{shen2018fractional} via a sequence of quadratic subproblems
given as
\begin{equation}
\underset{\mathbf{w}\in\mathcal{W}}{\mathsf{minimize}}\;\;-\frac{2\left(\mathbf{x}^{k}-r_{f}\right)}{\mathbf{y}^{k}}\left(x\left(\mathbf{w}\right)-r_{f}\right)+\left(\frac{\mathbf{x}^{k}-r_{f}}{\mathbf{y}^{k}}\right)^{2}y\left(\mathbf{w}\right).\label{eq:qt msrp subprob}
\end{equation}
Note that Problem (\ref{eq:qt msrp subprob}) is in the form of Problem
(\ref{eq:QP surrogate}), and the weight updates coincide with that
in SCQP after scaling. Hence, the quadratic transform is a special
case of SCQP.

\subsubsection*{On the connection to Dinkelbach\textquoteright s algorithm for MGSRP}

To solve the MGSRP problem whose objective is (\ref{eq:mgsr form})
given $\beta=1$, Dinkelbach's algorithm \cite{dinkelbach1967} deals
with the following surrogate problem
\begin{equation}
\underset{\mathbf{w}\in\mathcal{W}}{\mathsf{minimize}}\;\;-\left(x\left(\mathbf{w}\right)-r_{f}\right)+\left(\frac{\mathbf{x}^{k}-r_{f}}{\mathbf{y}^{k}}\right)y\left(\mathbf{w}\right),\label{eq:dinkelback mgsrp subprob}
\end{equation}
equivalent to the one in SCQP\@. Therefore, SCQP can be readily specialized
as Dinkelbach's algorithm.

\subsubsection*{On the connection to MM for worst-case robust GMRP}

To solve the worst-case robust GMRP problem whose objective is (\ref{eq:worst-case GMRP}),
the MM algorithm \cite{sun2016majorization} iteratively approximates
the $\mathcal{\ell}_{2}$-norm and constructs a sequence of upper-bound
problems
\begin{equation}
\underset{\mathbf{w}\in\mathcal{W}}{\mathsf{minimize}}\;\;-x\left(\mathbf{w}\right)+\frac{\alpha}{2}\left(\frac{y\left(\mathbf{w}\right)}{\sqrt{\mathbf{y}^{k}}}+\sqrt{\mathbf{y}^{k}}\right).\label{eq:mm GMRP}
\end{equation}
We notice that Problem (\ref{eq:mm GMRP}) matches Problem (\ref{eq:QP surrogate})
with the same updates of weights as SCQP, which means MM can be interpreted
as a special case of SCQP.

We remark again that the existing algorithms are only preferred in
specific cases when their surrogate problems are easy to compute.
In contrast, our algorithmic framework is more accessible because
it can always solve \ref{eq:General MVP formulation} via a series
of simple QP surrogate problems. We refer the reader to Sections \ref{sec:Analysis}
and \ref{sec:Fast} for more detailed analysis and design, and Section
\ref{sec:Applications-and-Experiments} for the performance in representative
applications.

\section{Analysis of Proposed Algorithm\label{sec:Analysis}}

In the previous section, we have offered an overview of the proposed
algorithm to solve \ref{eq:General MVP formulation}. In this section,
we provide more details and insights. In subsection \ref{subsec:Insight-from-Multiobjective},
we show how \ref{eq:General MVP formulation} and the proposed algorithm
strongly relate to Pareto optimality. In subsections \ref{subsec:Making-Objective-Function}
and \ref{subsec:Relaxing-Constraints-to}, we elaborate the technical
details of constructing QP surrogate problems, and in subsection \ref{subsec:Convergence-Analysis},
we provide the convergence analysis.

\subsection{Insight from Multiobjective Optimization\label{subsec:Insight-from-Multiobjective}}

We begin by interpreting general MVP optimization using the tools
from multiobjective optimization. Given multiple estimates of expected
return and risk in (\ref{eq:mean and variance terms}), portfolio
selection can be treated as a multiobjective optimization problem
\begin{equation}
\underset{\mathbf{w}\in\mathcal{W}}{\mathsf{minimize}}\;\;\{-x_{1}(\mathbf{w}),\dots,-x_{p}(\mathbf{w}),y_{1}(\mathbf{w}),\dots,y_{q}(\mathbf{w})\}.\label{eq:multiobj}
\end{equation}
Ideally, investors attempt to obtain the portfolio $\mathbf{w}\in\mathcal{W}$
that simultaneously maximizes all $x_{i}(\mathbf{w})$ and minimizes
all $y_{j}(\mathbf{w})$. However, such a portfolio does not usually
exist due to competing objectives. Therefore, Pareto optimal solutions,
a.k.a. efficient portfolios, are favored by investors. Given a Pareto
optimal solution, we cannot find any other solution that improves
one objective (i.e., higher return or lower risk) without degrading
at least one of the other objectives. The set of Pareto optimal solutions
constitutes the Pareto frontier. Although \ref{eq:General MVP formulation}
can represent various risk-return preferences by different formulations,
its design should obey a rational risk-return trade-off discipline
and produce a Pareto optimal solution. For ease of interpretation,
we denote the stationary solution of \ref{eq:General MVP formulation}
as $\mathbf{w}^{\star}$ and make the following assumption.

\begin{assumption} For each $x_{i}$ and $y_{j}$ that does not exist
in $F$, we have
\[
x_{i}\left(\mathbf{w}^{\star}\right)=a_{i},\quad y_{j}\left(\mathbf{w}^{\star}\right)=b_{j}.
\]

\label{ASMP: g touch}\end{assumption}

The above assumption requires that for all $x_{i}$ and $y_{j}$ that
does not exist in $F$, their corresponding mean-variance constraints
must be active at $\mathbf{w}^{\star}$. Otherwise, these estimates
can be safely removed from \ref{eq:General MVP formulation} without
affecting $\mathbf{w}^{\star}$. We note that this assumption is not
required by the proposed algorithm. Then, the connection between \ref{eq:General MVP formulation}
and Pareto optimality can be established as follows.
\begin{lem}
\label{lem:Pareto}Under Assumption \ref{ASMP: Pareto} and \ref{ASMP: g touch},
every stationary solution of \ref{eq:General MVP formulation} is
a Pareto optimal solution of Problem (\ref{eq:multiobj}).
\end{lem}
\begin{figure}
\begin{centering}
\includegraphics[width=0.93\columnwidth]{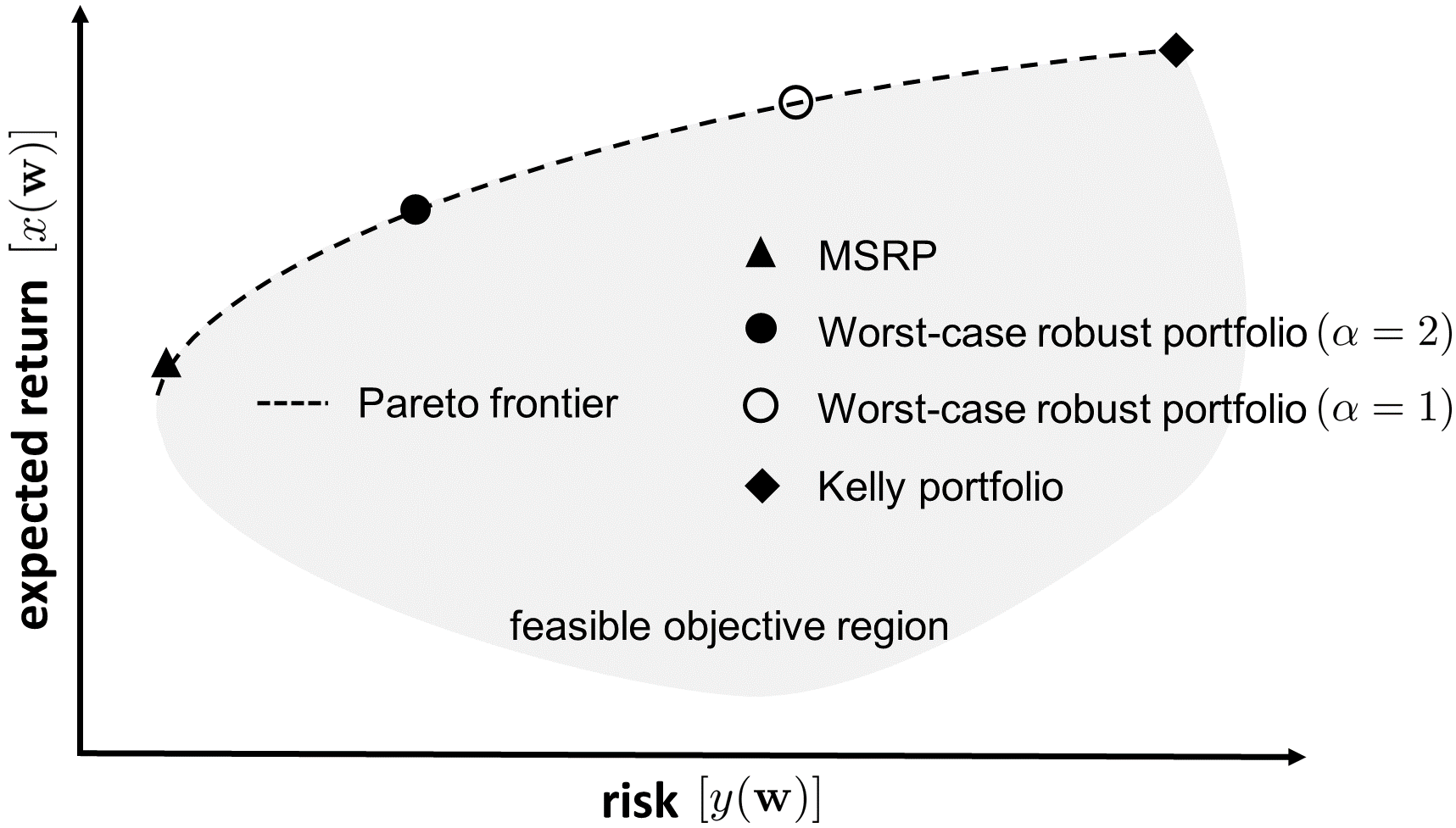}
\par\end{centering}
\caption{MVPs in the risk-return objective space for illustrative purposes.
\label{fig:Pareto frontier}}
\end{figure}
Figure \ref{fig:Pareto frontier} depicts the relationship between
several MVPs and the Pareto frontier. It is based on a bi-objective
optimization problem with a single expected return $x(\mathbf{w})$
and a single risk $y(\mathbf{w})$. The feasible objective region
in gray contains all possible objective vectors that could be achieved
by any $\mathbf{w}\in\mathcal{W}$. Therefore, its upper left boundary,
highlighted with a dashed curve, is the Pareto frontier. We observe
that all the exhibited portfolios reside along the Pareto frontier.
By changing the parameter $\alpha$ of the worst-case robust portfolio,
we can also achieve different Pareto optimal solutions.

We now show how the QP surrogate problem (\ref{eq:QP surrogate})
relates to the Pareto frontier. To characterize the Pareto optimal
solutions, the weighting method solves the following problem

\begin{equation}
\underset{\mathbf{\mathbf{w}\in\mathcal{W}}}{\mathsf{minimize}}\;\;-\mathbf{v}_{x}^{\intercal}\mathbf{x}\left(\mathbf{w}\right)+\mathbf{v}_{y}^{\intercal}\mathbf{y}\left(\mathbf{w}\right),\label{eq:multiobj-weighted sum}
\end{equation}
where $\mathbf{v}_{x}$ and $\mathbf{v}_{y}$ are non-negative coefficients.
The weighing coefficients have the physical meaning of reflecting
the relative importance of the objectives \cite{marler2010weighted}.
Previous studies have demonstrated that Problem (\ref{eq:multiobj-weighted sum})
has two properties: every Pareto optimal solution of (\ref{eq:multiobj})
can be found by (\ref{eq:multiobj-weighted sum}) given proper $\mathbf{v}_{x}$
and $\mathbf{v}_{y}$; the unique solution of (\ref{eq:multiobj-weighted sum})
is always Pareto optimal \cite{miettinen1999nonlinear}. In our case,
the surrogate problem (\ref{eq:QP surrogate}) coincides with Problem
(\ref{eq:multiobj-weighted sum}) when $\mathbf{v}_{x}$ and $\mathbf{v}_{y}$
are chosen as
\begin{equation}
\mathbf{v}_{x}=\boldsymbol{\lambda}_{x}+\boldsymbol{\eta}_{x},\quad\mathbf{v}_{y}=\boldsymbol{\lambda}_{y}+\boldsymbol{\eta}_{y}.
\end{equation}
Therefore, the surrogate problem (\ref{eq:QP surrogate}) inherits
the properties of Problem (\ref{eq:multiobj-weighted sum}). Based
on the above findings, we reveal our proposed algorithm's insight.
By dynamically adjusting $\boldsymbol{\lambda}$ and $\boldsymbol{\eta}$,
SCQP tracks the Pareto frontier as every point in its solution sequence
$\{\mathbf{\hat{w}}^{k}\}$ is Pareto optimal. When $\boldsymbol{\lambda}$
and $\boldsymbol{\eta}$ converge, the algorithm terminates at the
Pareto optimal solution that corresponds to the stationary solution
of \ref{eq:General MVP formulation}. This insight distinguishes our
proposed algorithm from the general state-of-the-art methods. As the
Pareto frontier has a number of attractive features, SCQP can benefit
from them and has the fast implementation given in Section \ref{sec:Fast}.

\begin{figure}
\begin{centering}
\includegraphics[width=0.9\columnwidth]{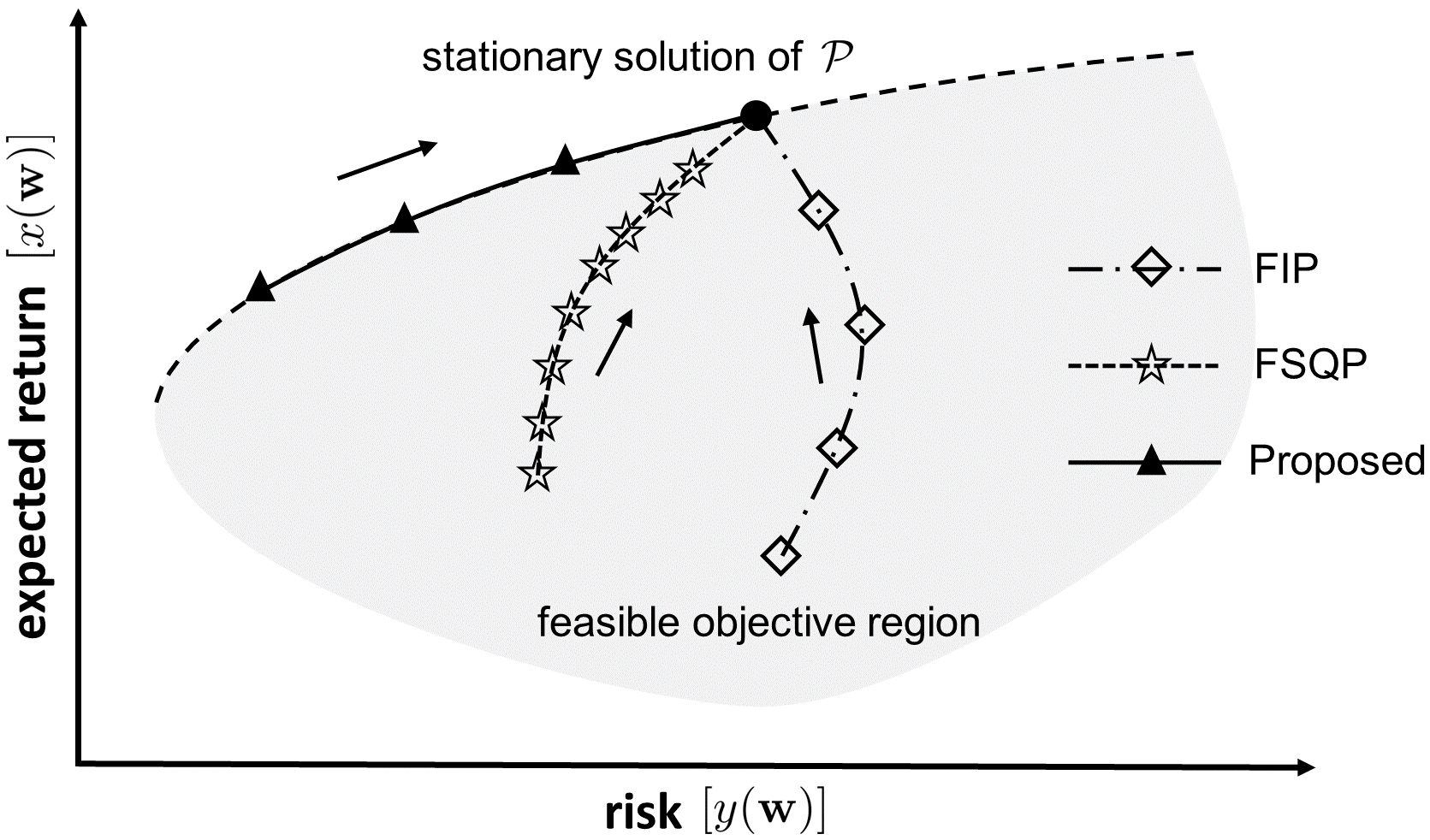}
\par\end{centering}
\caption{Iterates of FIP, FSQP, and SCQP plotted in the risk-return objective
space for illustrative purposes. \label{fig:solution-path}}
\end{figure}

Figure \ref{fig:solution-path} exhibits how iterates of FIP, FSQP,
and SCQP shift in the risk-return objective space when solving \ref{eq:General MVP formulation}.
Specifically, for FIP and FSQP whose subproblems have no guarantee
of Pareto optimal solutions, usually they do not achieve Pareto optimality
before termination.

\subsection{Outer Loop: Making Objective Function Quadratic\label{subsec:Making-Objective-Function}}

 In Section \ref{sec:Proposed-Algorithm}, we iteratively approximate
$f$ with the quadratic function $\tilde{f}$ in the form of the weighted
combination of means and variances. When the outer loop converges
to its fixed point, denoted by $\mathbf{w}^{\star}$ with a slight
abuse of notation, we obtain
\begin{equation}
\left(\mathbf{z}-\mathbf{w}^{\star}\right)^{\intercal}\nabla\tilde{f}\left(\mathbf{w}^{\star};\mathbf{w}^{\star}\right)\ge0,\:\text{for all }\mathbf{z}\in\mathcal{K},\label{eq:first-order condition of f}
\end{equation}
from the optimality conditions of Problem (\ref{eq:f app}). It is
easy to observe that $\mathbf{w}^{\star}$ is also a stationary solution
of \ref{eq:General MVP formulation} whose stationary condition matches
(\ref{eq:first-order condition of f}) as $\nabla f(\mathbf{w}^{\star})=\nabla\tilde{f}(\mathbf{w}^{\star};\mathbf{w}^{\star})$.
Therefore, the outer loop should converge only when it meets the stationary
solution of \ref{eq:General MVP formulation}. The rationality of
using successive quadratic approximation with $\tilde{f}$ to solve
\ref{eq:General MVP formulation} can be formally established in the
SCA framework \cite{scutari2016parallel,scutari2013decomposition}
when the following additional assumption is made.

\begin{assumption}At least one $y_{j}(\mathbf{w})$ exists in $F$.
In addition, $f$ has a Lipschitz continuous gradient on $\mathcal{K}$.

\label{ASMP: F contain y}\end{assumption}

Note that Assumption \ref{ASMP: F contain y} is quite standard and
is satisfied by a large class of practical problems. Therefore, $\tilde{f}$
has the following properties: for all $\mathbf{w}^{k}\in\mathcal{K}$,
\begin{itemize}
\item $\tilde{f}(\mathbf{w};\mathbf{w}^{k})$ is differentiable and $\nabla f(\mathbf{w}^{k})=\nabla\tilde{f}(\mathbf{w}^{k};\mathbf{w}^{k})$;
\item $\tilde{f}(\mathbf{w};\mathbf{w}^{k})$ is strongly convex on $\mathcal{K}$.
\end{itemize}
These properties confirm that $\tilde{f}$ is a suitable approximant
in the SCA framework, which requires consistency in the first-order
derivative of $f$ and strong convexity \cite{scutari2018parallel}.
$\boldsymbol{\lambda}$ computed as the partial derivatives in (\ref{eq:lambda update})
represents the intended preferences of $f$ over the objectives, aligning
with the interpretation of the weighing coefficients in Problem (\ref{eq:multiobj-weighted sum}).

We note that there are other approximants under the SCA framework.
They may have outstanding performance in risk parity portfolio optimization
\cite{feng2015scrip} and high-order portfolio optimization \cite{zhou2021sca}.
However, in our case, the proposed $\tilde{f}$ is preferred due to
the following advantages. First, it does not require a proximal-like
regularization term that adds to the strong convexity for convergence
as it is strongly convex in nature by taking a positive weighted sum
of $y_{j}(\mathbf{w})$. Second, it grasps the benefits of tracking
the Pareto frontier for a much more efficient implementation.

\subsection{Inner Loop: Dealing with Mean-variance Constraints\label{subsec:Relaxing-Constraints-to}}

When the mean-variance constraints exist in Problem (\ref{eq:f app}),
we apply the partial relaxation of $\mathbf{g}(\mathbf{w})\le\mathbf{0}$
and then obtain the partial Lagrangian written as
\begin{equation}
\begin{aligned}L(\mathbf{w},\boldsymbol{\eta};\boldsymbol{\lambda}^{k})= & \tilde{f}(\mathbf{w};\mathbf{w}^{k})-\boldsymbol{\eta}_{x}^{\intercal}\left(\mathbf{x}\left(\mathbf{w}\right)-\mathbf{a}\right)\\
 & +\boldsymbol{\eta}_{y}^{\intercal}\left(\mathbf{y}\left(\mathbf{w}\right)-\mathbf{b}\right),
\end{aligned}
\end{equation}
where $\boldsymbol{\eta}_{x}\in\mathbb{R}_{+}^{p}$ and $\boldsymbol{\eta}_{y}\in\mathbb{R}_{+}^{q}$
are the Lagrangian multipliers associated with the constraints $\mathbf{x}(\mathbf{w})\ge\mathbf{a}$
and $\mathbf{y}(\mathbf{w})\le\mathbf{b}$, respectively \cite{palomar2007alternative}.
The Lagrangian dual function $h$ is defined as the infimum of $L$
over $\mathbf{w}\in\mathcal{W}$:
\begin{equation}
h(\boldsymbol{\eta};\boldsymbol{\lambda}^{k})=\underset{\mathbf{w}\in\mathcal{W}}{\mathsf{inf}}\:L(\mathbf{w},\boldsymbol{\eta};\boldsymbol{\lambda}^{k}),\label{eq:partial dual function}
\end{equation}
where the infimum is achieved at $\tilde{\mathbf{w}}(\boldsymbol{\eta};\boldsymbol{\lambda}^{k})$. 

The strong duality holds as Problem (\ref{eq:f app}) is convex and
has at least one strictly feasible point \cite{boyd2004convex}. Therefore,
instead of solving the primal problem, we can optimize its dual problem
(\ref{eq:partial dual problem}) using the gradient projection method.
According to \cite{bertsekas1999nonlinear}, the gradient of $h$
can be computed as
\begin{equation}
\nabla h(\boldsymbol{\eta};\boldsymbol{\lambda}^{k})=\mathbf{g}(\tilde{\mathbf{w}}(\boldsymbol{\eta};\boldsymbol{\lambda}^{k})).
\end{equation}
An important feature is that the partial minimization problem in (\ref{eq:partial dual function})
coincides with our QP surrogate problem (\ref{eq:QP surrogate}),
i.e., we have $\tilde{\mathbf{w}}(\boldsymbol{\eta};\boldsymbol{\lambda}^{k})=\hat{\mathbf{w}}(\boldsymbol{\lambda}^{k},\boldsymbol{\eta})$.
In other words, each $\nabla h$ can be simply obtained from a QP
(\ref{eq:QP surrogate}) as shown in (\ref{eq:grad h with QP}).

The Lagrangian multiplier $\boldsymbol{\eta}$ has the following explanation.
Assume $\eta_{j}$ corresponds to the risk constraint $y_{j}(\mathbf{w})\le b_{j}$.
If this constraint is satisfied with strict inequality $y_{j}(\mathbf{w})<b_{j}$,
$\eta_{j}$ will decrease according to (\ref{eq:eta update}) and
(\ref{eq:grad h with QP}). From the view of the weighing method,
the weight on $y_{j}(\mathbf{w})$ will decrease, which means more
risk $y_{j}(\mathbf{w})$ is preferred in exchange for a better risk-return
trade-off in the next iteration. Otherwise, if the risk constraint
is violated with $y_{j}(\mathbf{w})>b_{j}$, $\eta_{j}$ will increase,
and thus the concern about the risk $y_{j}(\mathbf{w})$ will rise.
The above interpretation can also be applied to the expected return
constraint $x_{i}(\mathbf{w})\ge a_{i}$.

\subsection{Convergence Analysis\label{subsec:Convergence-Analysis}}

In this subsection, we provide the convergence of the proposed SCQP.
We note that the convergence of the outer loop is based on that of
the inner loop. 

To ensure the convergence of the inner loop, we choose $\alpha^{l}$
according to the Armijo rule along the projection arc \cite{bertsekas1999nonlinear}.
To be more specific, selecting $\alpha^{\text{init}}>0$, and $\sigma,\beta\in(0,1)$,
$\alpha^{l}$ is chosen to be the largest element in $\{\alpha^{\text{init}}\beta^{j}\}_{j=0,1,\dots}$
satisfying
\begin{equation}
h(\boldsymbol{\eta}^{l+1};\boldsymbol{\lambda}^{k})-h(\boldsymbol{\eta}^{l};\boldsymbol{\lambda}^{k})\ge\sigma\mathbf{g}(\hat{\mathbf{w}}(\boldsymbol{\lambda}^{k},\boldsymbol{\eta}^{l}))\strut^{\intercal}(\boldsymbol{\eta}^{l+1}-\boldsymbol{\eta}^{l}),\label{eq:Armijo-1}
\end{equation}
where
\begin{equation}
\boldsymbol{\eta}^{l+1}=\left[\boldsymbol{\eta}^{l}+\alpha^{l}\mathbf{g}(\hat{\mathbf{w}}(\boldsymbol{\lambda}^{k},\boldsymbol{\eta}^{l}))\right]_{+}.\label{eq:Armijo-2}
\end{equation}
Then, the convergence of the inner loop can be established as follows.
\begin{prop}
\label{prop:dual gradient converge}Under Assumption \ref{ASMP: Pareto},
suppose $\alpha^{l}$ is chosen according to Armijo rule along the
projection arc, then the sequence $\{\hat{\mathbf{w}}(\boldsymbol{\lambda}^{k},\boldsymbol{\eta}^{l})\}_{l=1}^{\infty}$
generated by the inner loop of Algorithm \ref{alg:successive QP}
converges to the optimal solution of Problem (\ref{eq:f app}).
\end{prop}
\begin{IEEEproof}
The convergence of the gradient projection method follows directly
from \cite[Proposition 2.3.3]{bertsekas1999nonlinear}. It states
that every limit point of $\{\boldsymbol{\eta}^{l}\}_{l=1}^{\infty}$
is a stationary solution of the dual problem (\ref{eq:partial dual problem}).
Given the strong duality holds, the generated sequence $\{\hat{\mathbf{w}}(\boldsymbol{\lambda}^{k},\boldsymbol{\eta}^{l})\}_{l=1}^{\infty}$
converges to the optimal solution of Problem (\ref{eq:f app}).
\end{IEEEproof}
Based on the convergence of the inner loop, we next analyze the convergence
of the outer loop.
\begin{prop}
\label{prop:convergence successive QP}Under Assumption \ref{ASMP: Pareto}
and \ref{ASMP: F contain y}, suppose $\gamma^{k}\in(0,1]$, $\gamma^{k}\rightarrow0$
and $\sum_{k}\gamma^{k}=+\infty$. Then, either Algorithm \ref{alg:successive QP}
converges in a finite number of iterations to a stationary solution
of \ref{eq:General MVP formulation} or every limit point of the solution
sequence $\{\mathbf{w}^{k}\}_{k=1}^{\infty}$ (at least one such point
exists) is a stationary solution of \ref{eq:General MVP formulation}.
\end{prop}
\begin{IEEEproof}
Note that problem \ref{eq:General MVP formulation} has a feasible
region $\mathcal{K}$ that is closed, bounded, convex and nonempty.
Thus, \cite[Assumptions A1-A4]{scutari2013decomposition} hold, and
the proof of Proposition \ref{prop:convergence successive QP} follows
directly from \cite[Theorem 3]{scutari2013decomposition}.
\end{IEEEproof}
Thus, by proving Proposition \ref{prop:dual gradient converge} and
\ref{prop:convergence successive QP}, we prove the convergence of
the proposed algorithm.

\section{Fast Implementation\label{sec:Fast}}

Since the updates of $\boldsymbol{\lambda}$ and $\boldsymbol{\eta}$
are generally cheap as mentioned, the major cost of the proposed algorithm
comes from solving the sequence of QP surrogate problems (\ref{eq:QP surrogate}).
Although open-source QP solvers are well established and considered
efficient for a single problem, solving a sequence of QP problems
is not a small workload. Therefore, it is necessary and attractive
to explore the hidden structures of SCQP for potential significant
speed-up. In this section, we first introduce the sparsity pattern
related to the Pareto frontier, and then propose a novel active-set
strategy to take advantage of this sparsity pattern for a fast implementation
of SCQP.

\subsection{Sparsity Pattern on Pareto Frontier}

One of the crucial features of the Pareto frontier is the sparsity
pattern. Previous studies have shown that Pareto optimal solutions
are naturally sparse under the long-only constraint \cite{brodie2009sparse,tian2019sparse}.
One explanation is that the long-only constraint acts like the $\ell_{1}$-norm
penalty which is extensively used to promote sparsity \cite{kondor2017analytic}. 

\begin{figure}
\begin{centering}
\includegraphics[width=0.95\columnwidth]{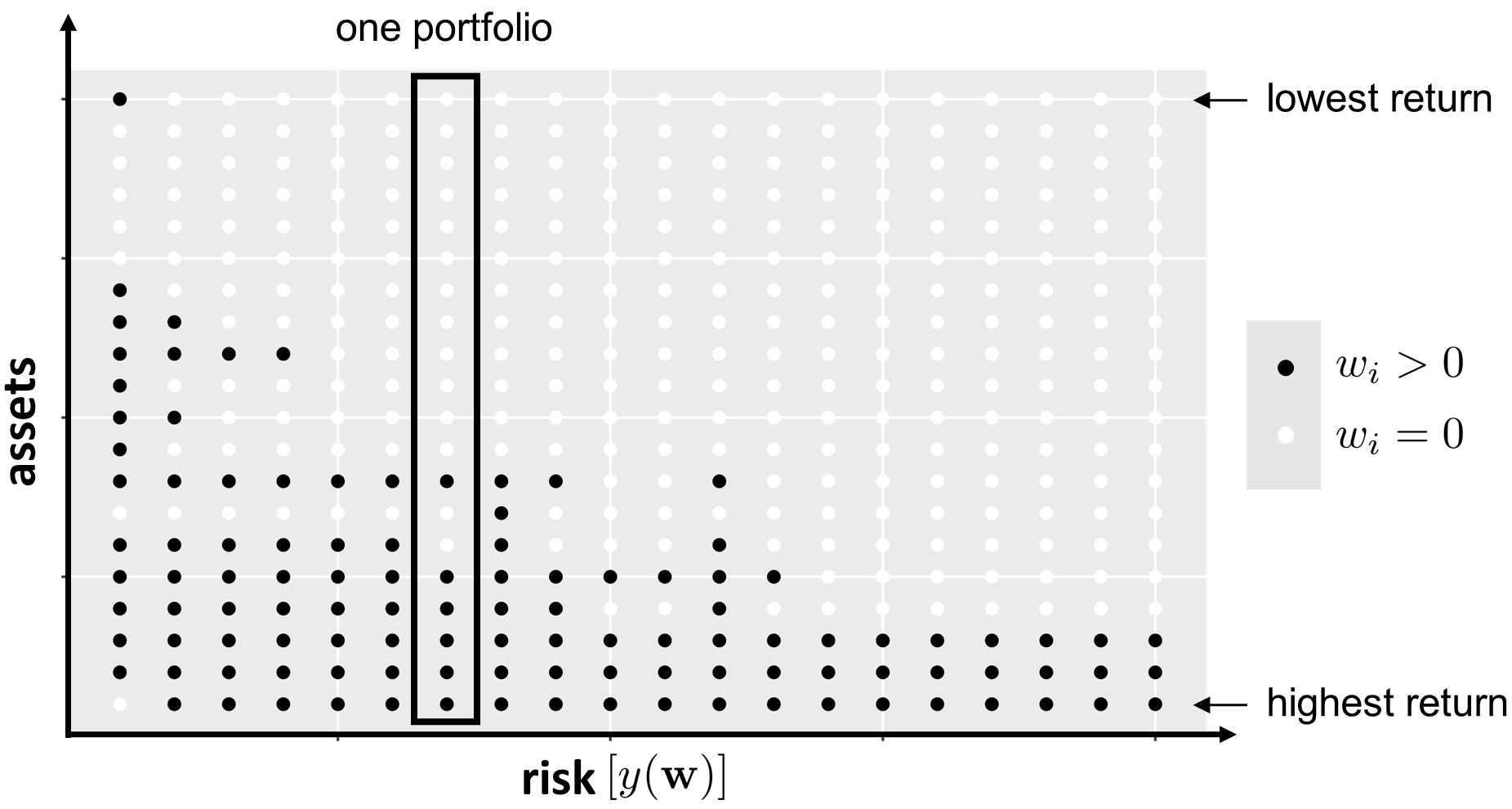}
\par\end{centering}
\caption{Sparsity pattern in efficient portfolios. \label{fig:sparsity-pattern}}
\end{figure}
To better illustrate the sparsity pattern, Figure \ref{fig:sparsity-pattern}
visualizes the efficient portfolios of $N=20$ given different levels
of risk aversion, which is based on the trade-off between a single
expected return $x(\mathbf{w})$ and a single risk $y(\mathbf{w})$
in the S\&P 500 market. Two observations from Figure \ref{fig:sparsity-pattern}
should be noted. First, the sparsity is remarkable, and the portfolios
are concentrated on a small number of assets with high returns, especially
when the risk tolerance is high. Second, the sparsity pattern between
any two neighboring portfolios is similar.

The above two observations have important algorithmic implications.
The first observation motivates us to solve each QP surrogate problem
(\ref{eq:QP surrogate}) via an equivalent dimension-reduced problem
defined on only a small portion of assets corresponding to $w_{i}>0$.
Because the complexity of the QP solver is around $\mathcal{O}(N^{3})$
\cite{ye1989extension,kang2020large}, solving the dimension-reduced
problem can be computationally much cheaper than directly handling
the original problem, especially when $N$ is large. The second observation
inspires that the solution of the previous QP can be used to \textquotedblleft warm
start\textquotedblright{} the solution of the next so that QP algorithms
benefit from the similar sparsity pattern and thus have faster convergence.
The following subsection details how we achieve the above speedup.

\subsection{New Active-set Strategy for QP}

We focus on the general form of the QP surrogate problem (\ref{eq:QP surrogate})
given as
\begin{equation}
\begin{aligned}\underset{\mathbf{w}\in\mathbb{R}^{N}}{\mathsf{minimize}}\,\,\, & \quad q\left(\mathbf{w}\right)=\mathbf{c}^{\intercal}\mathbf{w}+\frac{1}{2}\mathbf{w}^{\intercal}\mathbf{H}\mathbf{w}\\
\mathsf{subject\,\,to} & \quad\mathbf{A}\mathbf{w}=\mathbf{b},\quad\mathbf{l}\le\mathbf{w}\le\mathbf{u},
\end{aligned}
\label{eq:general QP}
\end{equation}
where $\mathbf{H}\in\mathbb{S}_{++}^{N}$, $\mathbf{A}\in\mathbb{R}^{M\times N}$,
$\mathbf{b}\in\mathbb{R}^{M}$, and $\mathbf{l},\mathbf{u}\in\mathbb{R}^{N}$
that specify bounds on the variables. Specifically, suppose $\mathbf{A}=\mathbf{1}\strut^{\intercal}$,
$\mathbf{b}=1$, $\mathbf{l}=\mathbf{0}$, and $\mathbf{u}=+\infty$,
the set of constraints in Problem (\ref{eq:general QP}) specializes
to $\mathcal{W}$. Empirically, a large proportion of the variables
$w_{i}$ will touch the boundaries, i.e., $w_{i}=l_{i}$ or $w_{i}=u_{i}$,
when they meet the optimality, which results in sparsity in our case.
Hence, if we know such variables in advance, we can instead search
for the solution of Problem (\ref{eq:general QP}) on a subspace spanned
by the remaining variables or, equivalently, solve the following small-sized
problem
\begin{equation}
\begin{aligned}\underset{\mathbf{w}\in\mathbb{R}^{N}}{\mathsf{minimize}}\,\,\, & \quad\mathbf{c}^{\intercal}\mathbf{w}+\frac{1}{2}\mathbf{w}^{\intercal}\mathbf{H}\mathbf{w}\\
\mathsf{subject\,\,to} & \quad\mathbf{A}\mathbf{w}=\mathbf{b},\quad l_{i}\le w_{i}\le u_{i},\:i\notin\bar{L}\cup\bar{U},\\
 & \quad w_{i}=l_{i},\:i\in\bar{L},\quad w_{i}=u_{i},\:i\in\bar{U},
\end{aligned}
\label{eq:ASM_sub}
\end{equation}
where $\bar{L}\cup\bar{U}$ is called a working set, and $\bar{L}\cap\bar{U}=\emptyset$.
Problem (\ref{eq:ASM_sub}) is generally low-cost because it can be
reduced to an $N-|\bar{L}\cup\bar{U}|$ dimensional QP by eliminating
the variables fixed on the bound. By analyzing the difference in optimality
conditions of Problem (\ref{eq:general QP}) and Problem (\ref{eq:ASM_sub}),
solving the reduced problem is equivalent to solving the original
problem if and only if the following conditions are satisfied
\begin{equation}
\beta_{i}^{l}\ge0,\:\forall i\in\bar{L},\quad\beta_{i}^{u}\ge0,\:\forall i\in\bar{U},\label{eq:asm optimality}
\end{equation}
 where $\beta_{i}^{l}$ and $\beta_{i}^{u}$ are the Lagrangian multipliers
corresponding to $w_{i}=l_{i}$ and $w_{i}=u_{i}$.

The above findings further inspire our new active-set strategy in
two aspects. First, we hope that the size of the working set can be
as large as possible for the sake of computational efficiency. Second,
if the current subproblem (\ref{eq:ASM_sub}) cannot solve (\ref{eq:general QP}),
the violation of condition (\ref{eq:asm optimality}) instructs the
update of the working set. The detailed procedure is given as follows. 

\begin{algorithm}
\caption{New Active-set Strategy for Problem (\ref{eq:general QP}). \label{alg:Active-set-strategy}}

\textbf{Input:} $\bar{L}^{0}$ and $\bar{U}^{0}$;

\begin{algorithmic}[1]

\For{$k=0,1,2,\ldots$}

\State Compute $\mathbf{w}^{k}$, $\boldsymbol{\beta}^{l}$, and
$\boldsymbol{\beta}^{u}$ via (\ref{eq:ASM_sub}) given $\{\bar{L}^{k},\bar{U}^{k}\}$;

\If {$\mathsf{min}(\boldsymbol{\beta}^{l})<0$ or $\mathsf{min}(\boldsymbol{\beta}^{u})<0$
}

\State $\bar{L}^{k+1}=\bar{L}^{k}\setminus\{i\left|\beta_{i}^{l}<0\right.\}$;

\State $\bar{U}^{k+1}=\bar{U}^{k}\setminus\{i\left|\beta_{i}^{u}<0\right.\}$;

\Else

\State Stop;

\EndIf

\EndFor

\end{algorithmic}

\textbf{Output:} The optimal solution $\mathbf{w}^{k}$ of Problem
(\ref{eq:general QP}).
\end{algorithm}
We begin with a feasible subproblem (\ref{eq:ASM_sub}) with a large
working set $\bar{L}^{0}\cup\bar{U}^{0}$. At the $k$th iteration,
we compute $\mathbf{w}^{k}$ and $(\boldsymbol{\beta}^{l},\boldsymbol{\beta}^{u})$
as the primal and dual optimal solutions of the subproblem (\ref{eq:ASM_sub})
based on the current working set $\bar{L}^{k}\cup\bar{U}^{k}$. If
condition (\ref{eq:asm optimality}) is satisfied, the algorithm terminates
as $\mathbf{w}^{k}$ meets the optimality conditions of Problem (\ref{eq:general QP}).
Otherwise, the objective function $q(\mathbf{w})$ may be further
decreased by relaxing the bounded variables corresponding to negative
multipliers. Hence, the working set for the subsequent iteration should
be adjusted as $\bar{L}^{k+1}=\bar{L}^{k}\setminus\{i\left|\beta_{i}^{l}<0\right.\}$
and $\bar{U}^{k+1}=\bar{U}^{k}\setminus\{i\left|\beta_{i}^{u}<0\right.\}$.
We formally describe this new active-set strategy in Algorithm \ref{alg:Active-set-strategy}.
Its convergence is summarized in the following proposition.
\begin{prop}
\label{prop:active set strategy}The sequence $\{\mathbf{w}^{k}\}$
generated by Algorithm \ref{alg:Active-set-strategy} converges to
the optimal solution of Problem (\ref{eq:general QP}) within $|\bar{L}^{0}\cup\bar{U}^{0}|$
iterations. 
\end{prop}
\begin{IEEEproof}
See Appendix \ref{apx: proof active set}.
\end{IEEEproof}

\subsubsection*{Warm starting for a sequence of QP}

In practice, Problem (\ref{eq:general QP}) can be solved in much
fewer iterations than the theoretical result in Proposition \ref{prop:active set strategy}.
When dealing with a sequence of related QPs, the number of iterations
can be further reduced using warm starting. This is because the optimal
working set stays mostly the same from one QP to the next because
of similar data input, manifesting as the similar sparsity pattern
in our case. To take advantage of this prior knowledge, the new active-set
strategy applies the warm starting that uses the optimal working set
of the former QP as an initial guess. In numerical experiments, Algorithm
\ref{alg:Active-set-strategy} with the warm starting typically solves
each QP surrogate problem (\ref{eq:QP surrogate}) in less than three
iterations, significantly reducing the cost of SCQP. 

\section{Numerical Experiments\label{sec:Applications-and-Experiments}}

We evaluate the applicability and performance of SCQP by conducting
experiments on representative MVPs.

\subsection{Experiment Set-Up}

\subsubsection{Real market data}

To evaluate the performance of SCQP, we perform experiments on historical
daily price time series data. Each dataset contains $N$ stocks randomly
chosen from the S\&P 500 index, and a time period of $5N$ continuous
trading days is randomly picked over the long period from 2008-12-01
to 2018-12-01. All results are averaged over $20$ independent realizations.

\subsubsection{Benchmarks}

The benchmarks for different MVPs are usually different. Specific
problems can be solved either individually or jointly by the following
methods:
\begin{itemize}
\item $\mathsf{ECOS}$: an open-source conic optimization software using
the interior point method \cite{domahidi2013ecos}; efficient for
small and medium-sized SOCP problems.
\item $\mathsf{MOSEK}$: a commercial conic optimization software using
the interior point method \cite{andersen2000mosek}; performs closely
to other commercial solvers like $\mathsf{GUROBI}$ and $\mathsf{CPLEX}$;
efficient for large-scale SOCP problems.
\item Dinkelbach's algorithm \& quadratic transform: iterative methods for
FP problems proposed in \cite{dinkelbach1967,shen2018fractional}.
\item Majorization-minimization (MM): an iterative method that solves difficult
optimization problems by solving a series of upper-bound surrogate
problems \cite{sun2016majorization,scutari2018parallel}.
\item $\mathsf{NLopt}$: a standard library for nonlinear optimization \cite{NLoptDocs};
applicable to general problems.
\item $\mathsf{DEoptim}$ and $\mathsf{GA}$: libraries of differential
evolution and genetic algorithm \cite{mullen2011deoptim,scrucca2013ga};
applicable to general problems.
\end{itemize}
Note that all the benchmarks safeguard the convergence, and thus our
comparisons focus on their convergence speed. It is generally difficult
to have a unified termination criterion for all methods as interior-point
solvers focus on the tolerance of primal and dual feasibility while
the other methods do not. Therefore, we set the termination criterion
of $\mathsf{ECOS}$ and $\mathsf{MOSEK}$ to their default options,
and terminate the other methods when
\begin{equation}
|\mathbf{w}^{k+1}-\mathbf{w}^{k}|\le10^{-6}.\label{eq:termination criteria}
\end{equation}

\subsubsection{Implementation details}

All experiments were carried out on 3.40GHz Intel Xeon Gold 6246R
machines with 80G RAM running R 3.6.3. The inner QP solver of SCQP
is the open-source $\mathsf{quadprog}$ \cite{goldfarb1983quadprog}.
In addition, we allow $\mathsf{DEoptim}$ and $\mathsf{GA}$ to run
in parallel mode with the same initial population of size 2000 as
the starting population, while the other methods run on only one core.

\subsection{Application I: Worst-case Robust GMRP\label{subsec:Application-I}}

\begin{table*}
\caption{Empirical time complexity order. \label{tab:empirical order}}

\centering{}%
\begin{tabular}{|c|c|}
\hline 
Application & Empirical time complexity order $\mathcal{O}\left(N^{c}\right)$\tabularnewline
\hline 
I: (\ref{eq:GMRP}) & Proposed (1.155) $<$ MOSEK (1.691) $<$ ECOS (2.453) $<$ NLopt (3.134)\tabularnewline
\hline 
II: (\ref{eq:Kelly}) & Proposed (0.944) $<$ MM (1.620) $<$ NLopt (3.241)\tabularnewline
\hline 
III: (\ref{eq:risk-const Markowitz}) & Proposed (1.137) $<$ MOSEK (1.615) $<$ ECOS (2.365) $<$ NLopt (3.140)\tabularnewline
\hline 
IV: (\ref{eq:risk-constrained MSRP}) & Proposed (0.768) $<$ Dinkelbach (1.786) $<$ QT (1.853) $<$ NLopt
(3.085)\tabularnewline
\hline 
\end{tabular}
\end{table*}
\begin{figure*}
\begin{centering}
\hspace{1em}\subfloat[Convergence of algorithms in a realization with $N=200$. \label{fig:robust convergence}]{\includegraphics[scale=0.3]{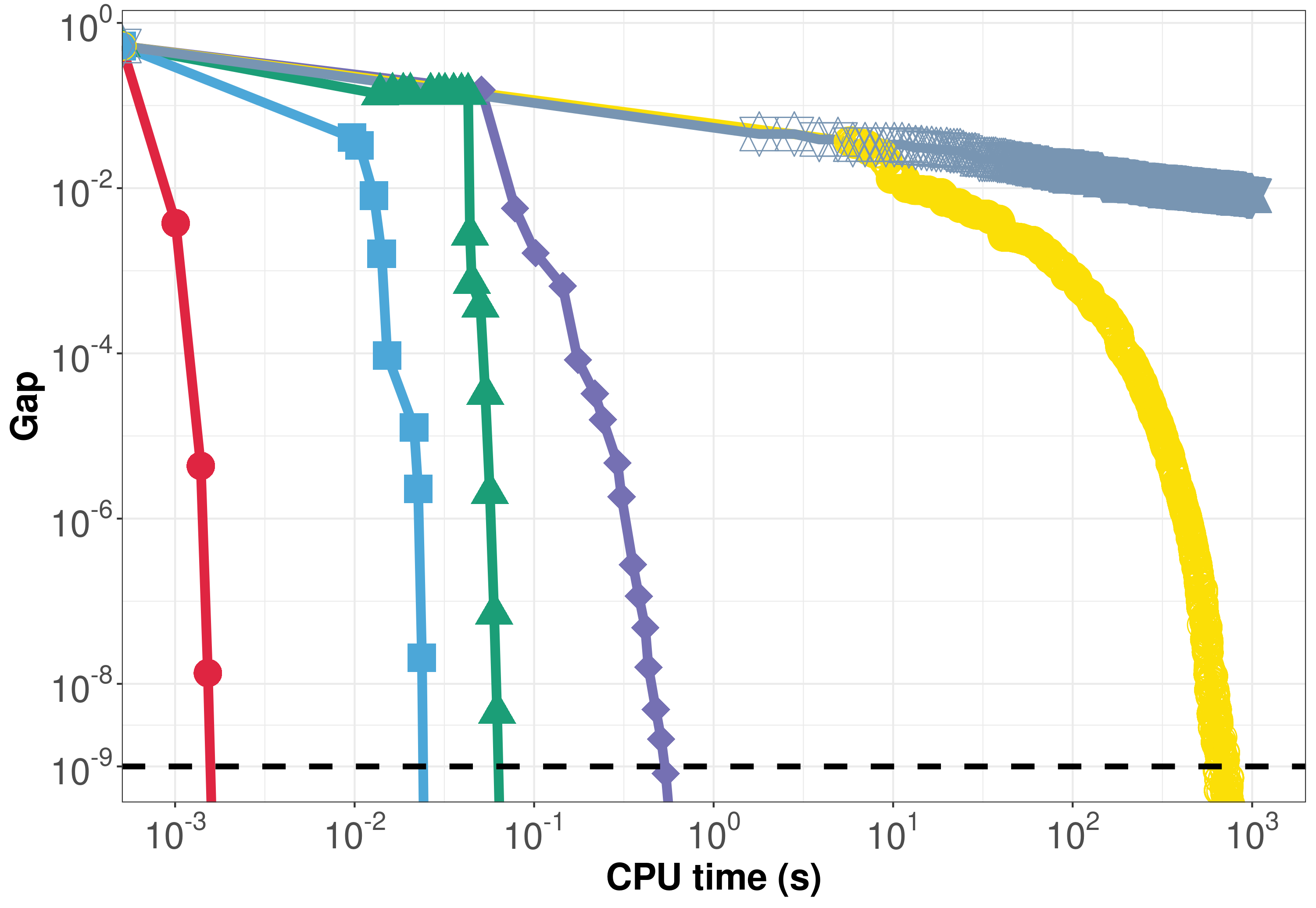}

}\hspace{2em}\subfloat[Timing results of algorithms with different $N$. \label{fig:robust dimension}]{\includegraphics[scale=0.3]{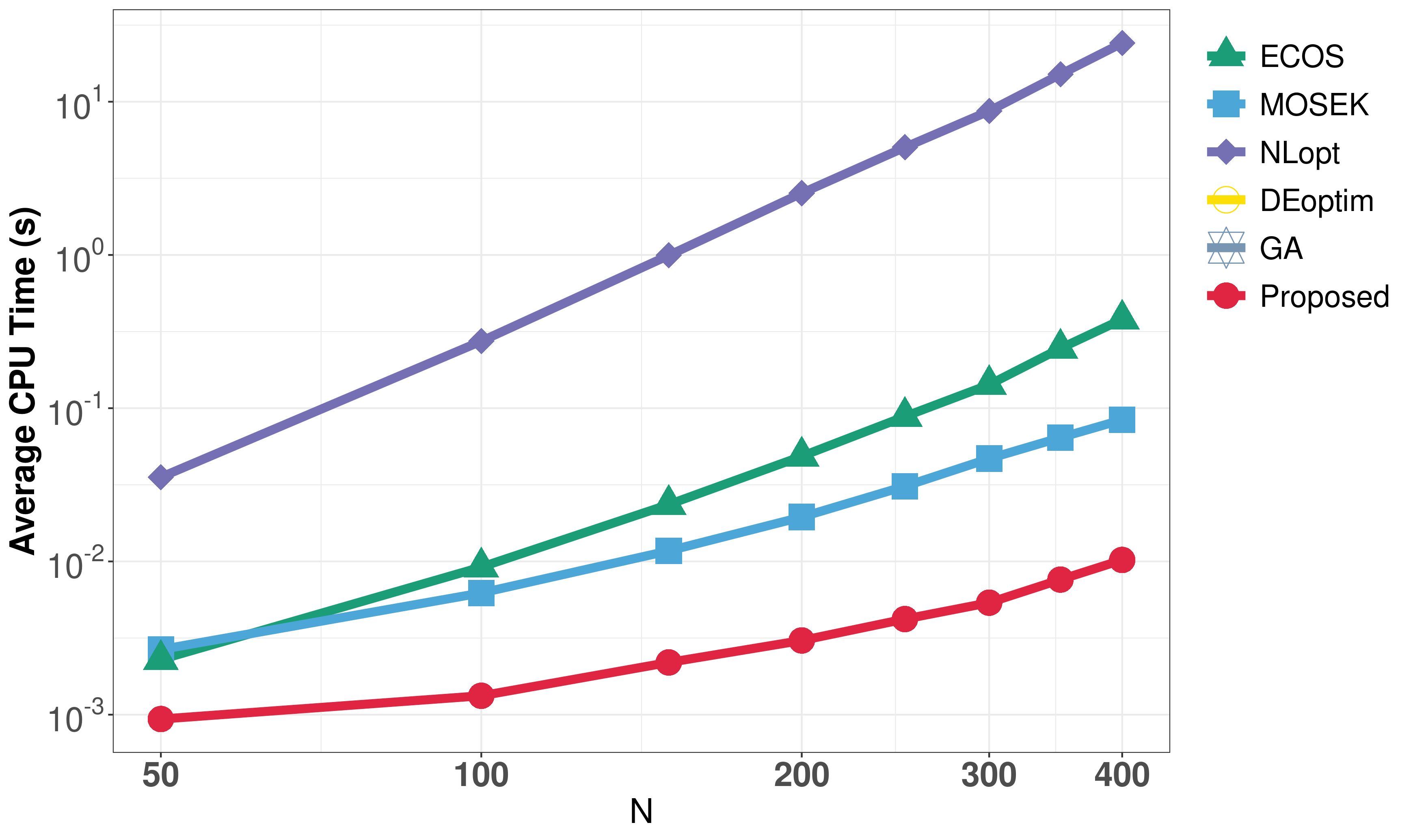}

}
\par\end{centering}
\caption{Numerical experiments on solving worst-case robust GMRP (\ref{eq:GMRP}).}
\end{figure*}
We consider the worst-case robust GMRP formulated as
\begin{equation}
\underset{\mathbf{w}\in\mathcal{W}}{\mathsf{minimize}}\;\;-\mathbf{w}^{\intercal}\boldsymbol{\mu}+\alpha\sqrt{\mathbf{w}\mystrut^{\intercal}\boldsymbol{\Sigma}\mathbf{w}},\label{eq:GMRP}
\end{equation}
where $\alpha>0$ is the predefined parameter. Without loss of generality,
we consider $\alpha=1$. Problem (\ref{eq:GMRP}) does not include
the mean-variance constraint $\mathbf{g}\left(\mathbf{w}\right)\le\mathbf{0}$,
so the inner loop of SCQP is not required. It is suitable for evaluating
the efficiency of the outer loop in dealing with convex objective
functions. Problem (\ref{eq:GMRP}) can be converted in the standard
way into a SOCP. Thus, SOCP solvers ($\mathsf{ECOS}$ and $\mathsf{MOSEK}$),
$\mathsf{NLopt}$, $\mathsf{DEoptim}$, and $\mathsf{GA}$ serve as
benchmarks.

Figure \ref{fig:robust convergence} shows the convergence of different
methods in one realization of Problem (\ref{eq:risk-constrained MSRP})
given $N=200$. The gap is defined as the absolute difference between
the objective value and the minimum one obtained by all the methods.
The plot shows that the proposed SCQP takes substantially less time
than the other methods to reach the gap of $10^{-9}$. We also observe
that SCQP consumes fewer iterations. Besides, $\mathsf{ECOS}$ and
$\mathsf{MOSEK}$ converge with similar performance as they share
similar interior point methods. As for the metaheuristic methods,
$\mathsf{DEoptim}$ converges to an accurate solution but has a much
slower rate, while $\mathsf{GA}$ cannot obtain good solutions within
1000s.

Figure \ref{fig:robust dimension} compares the CPU time of different
methods across different problem sizes $N$. $\mathsf{DEoptim}$ and
$\mathsf{GA}$ are omitted because they are time-consuming in this
application. The results show that SCQP consistently outperforms all
the other established solvers concerning computation time. More specifically,
when $N=50$, the proposed algorithm is $\sim2.5$x faster than $\mathsf{ECOS}$;
when $N\ge100$, the proposed algorithm has a $4.7\sim8.7$x faster
speed than $\mathsf{MOSEK}$. Another observation is that $\mathsf{ECOS}$
performs better than $\mathsf{MOSEK}$ in small problems ($N=50$)
and worse when the problem dimension becomes larger, which is also
verified by \cite{domahidi2013ecos}.

Table \ref{tab:empirical order} presents the empirical time complexity
order of different methods using $\mathcal{O}(N^{c})$ notation. They
are obtained by fitting the average CPU time vs $N$ curves (in a
log-log scale) with linear functions. We observe that SCQP has the
lowest empirical time complexity which is nearly linear. Besides,
$\mathsf{NLopt}$ has an empirical time complexity around $\mathcal{O}(N^{3})$,
which coincides with the discussion in \cite{NLoptDocs}. Combined
with Figure \ref{fig:robust dimension}, these results show that SCQP
is more scalable than the benchmarks. The major reason is that the
sparsity pattern becomes prominent in high-dimensional problems, while
SCQP benefits from it using the new active-set strategy for low computational
cost.

\subsection{Application II: Kelly Portfolio\label{subsec:Application-II}}

\begin{figure*}
\begin{centering}
\hspace{1em}\subfloat[Convergence of algorithms in a realization with $N=200$. \label{fig:kelly convergence}]{\includegraphics[scale=0.3]{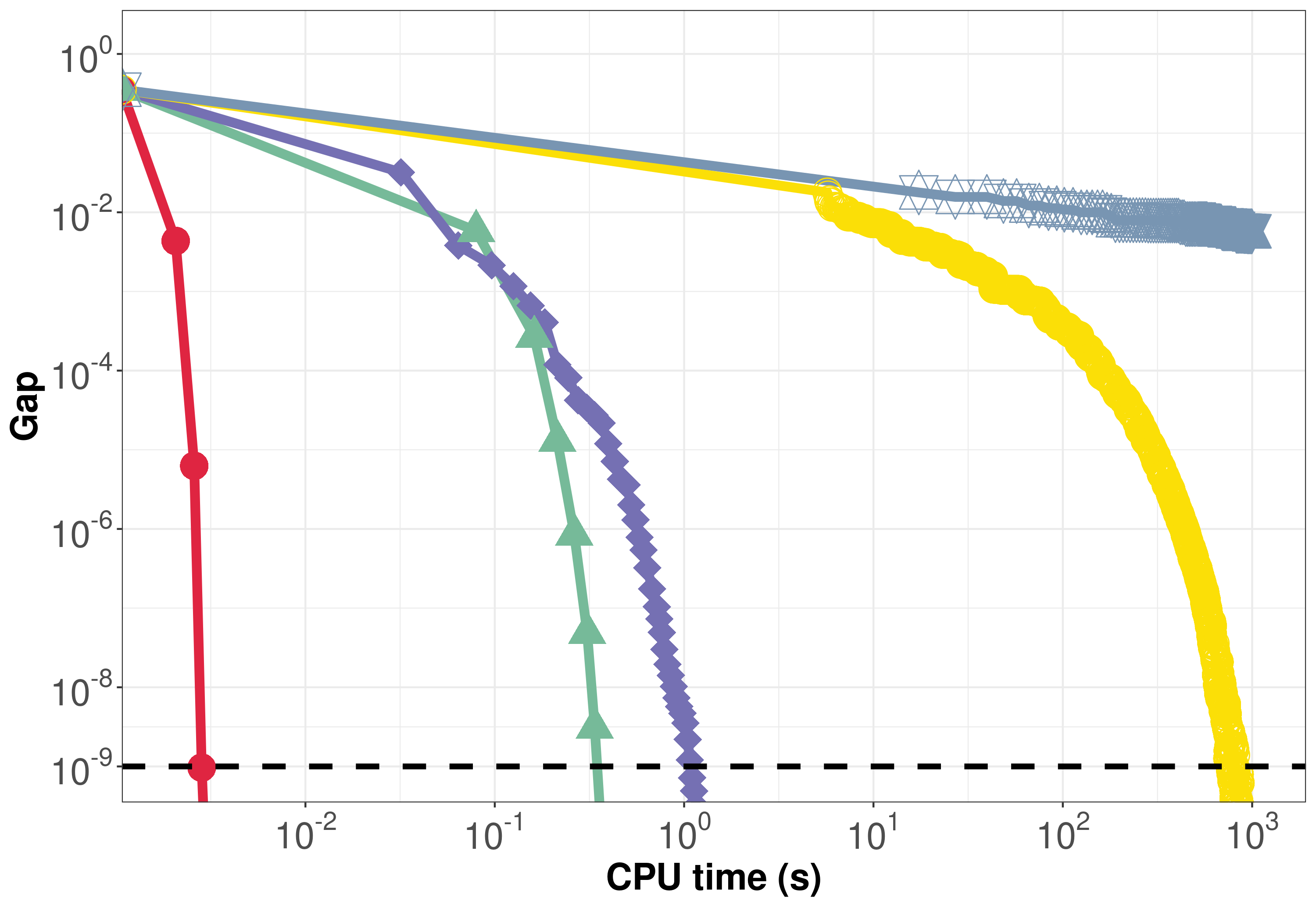}

}\hspace{2em}\subfloat[Timing results of algorithms with different $N$. \label{fig:kelly dimension}]{\includegraphics[scale=0.3]{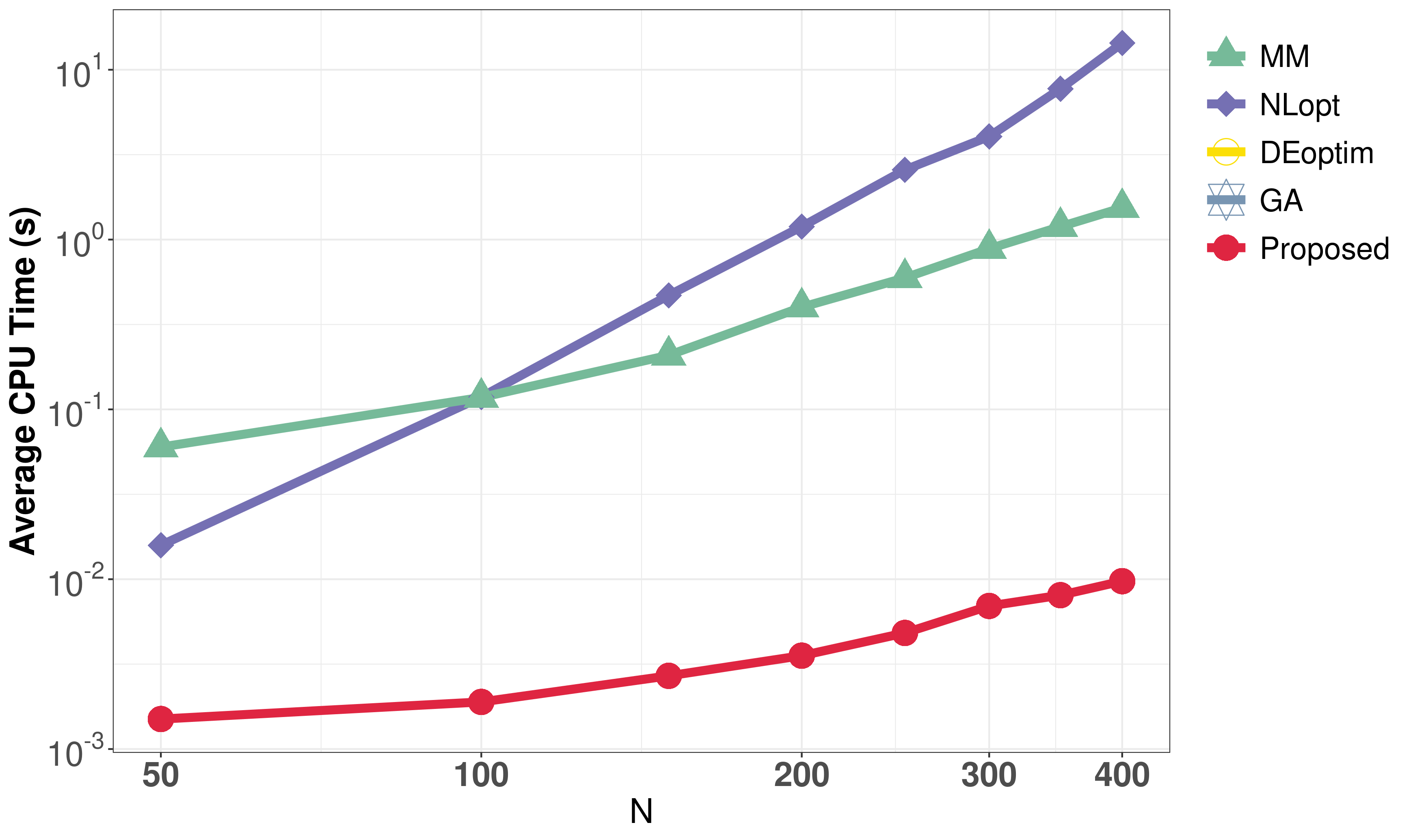}

}
\par\end{centering}
\caption{Numerical experiments on solving Kelly portfolio (\ref{eq:Kelly}).}
\end{figure*}
\begin{figure*}[t]
\begin{centering}
\hspace{1em}\subfloat[Convergence of algorithms in a realization with $N=200$. \label{fig:markowitz convergence}]{\includegraphics[scale=0.3]{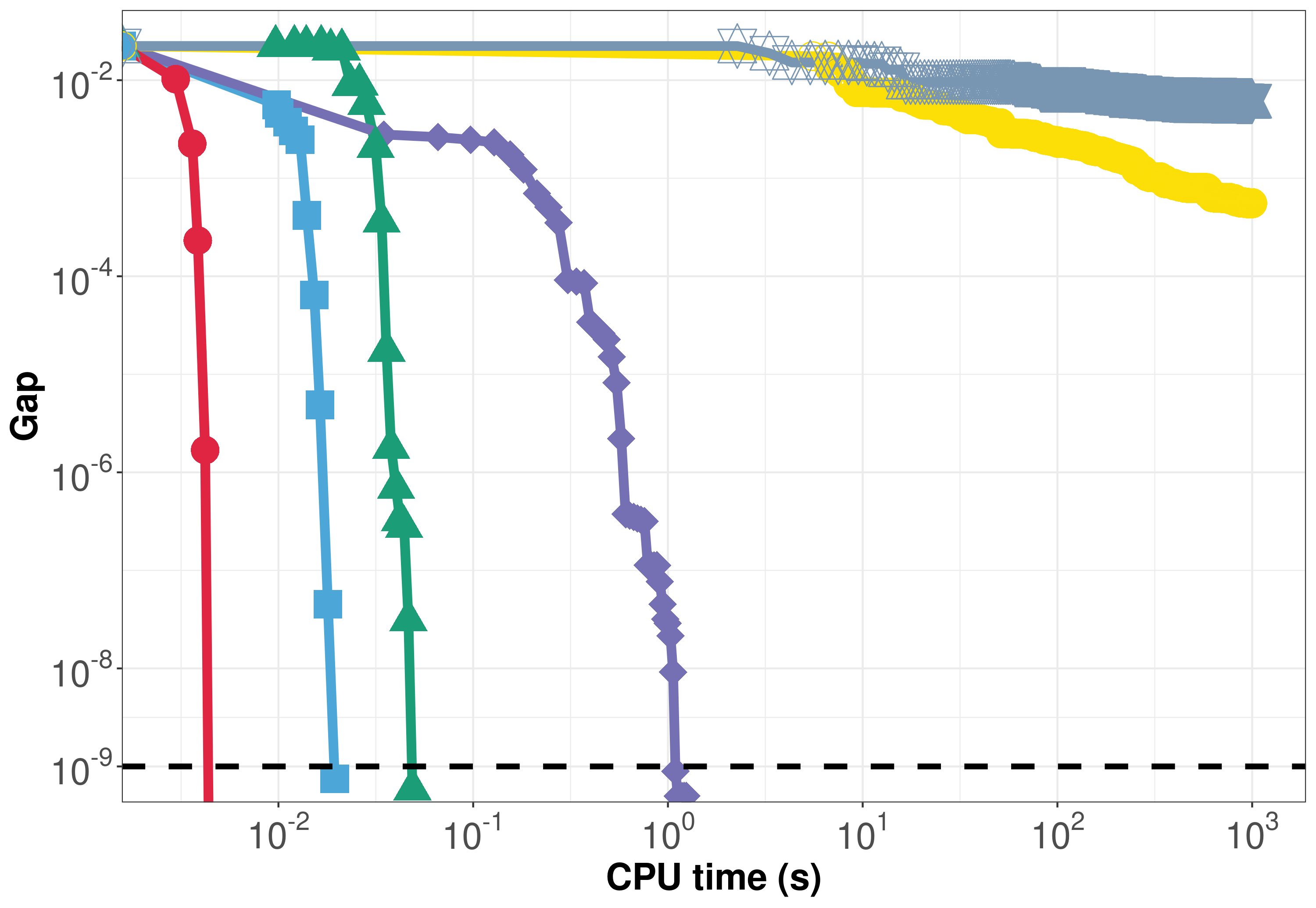}

}\hspace{2em}\subfloat[Timing results of algorithms with different $N$. \label{fig:markowitz dimension}]{\includegraphics[scale=0.3]{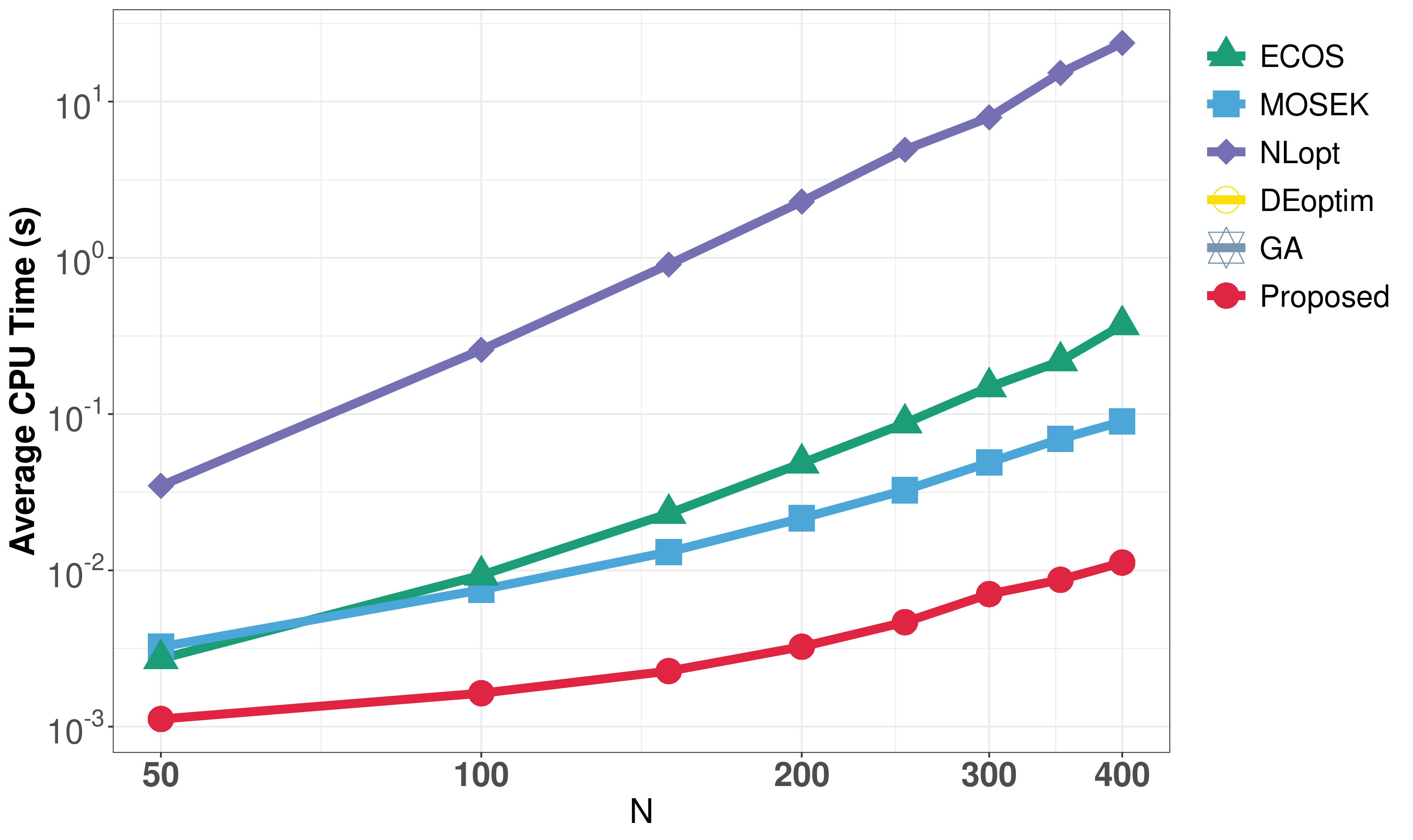}

}
\par\end{centering}
\caption{Numerical experiments on solving risk-constrained Markowitz portfolio
(\ref{eq:risk-const Markowitz}).}
\end{figure*}
In the first application, we show that the outer loop efficiently
deals with convex objective functions. Next, we focus on an MVP with
a nonconvex objective function. Recall the Kelly portfolio formulated
as
\begin{equation}
\underset{\mathbf{w}\in\mathcal{W}}{\mathsf{minimize}}\;\;-\log\left(1+\mathbf{w}^{\intercal}\boldsymbol{\mu}\right)+\frac{1}{2}\frac{\mathbf{w}^{\intercal}\boldsymbol{\Sigma}\mathbf{w}}{\left(1+\mathbf{w}\mystrut^{\intercal}\boldsymbol{\mu}\right)^{2}}.\label{eq:Kelly}
\end{equation}
The challenge of solving Problem (\ref{eq:Kelly}) is that it is a
nonconvex problem with a ratio term in the objective function. Similar
to Problem (\ref{eq:GMRP}), the inner loop of SCQP is also not required.

It is difficult to apply Dinkelbach's algorithm as Problem (\ref{eq:Kelly})
is not a traditional single-ratio FP. Alternatively, we introduce
the MM method that recasts the problem to a sequence of convex conic
subproblems in Appendix \ref{apx: MM for Kelly}. In addition, $\mathsf{NLopt}$,
$\mathsf{DEoptim}$, and $\mathsf{GA}$ are also applicable. Figure
\ref{fig:kelly convergence} shows the convergence of different methods
in one realization of Problem (\ref{eq:Kelly}) given $N=200$. The
proposed SCQP converges to the gap of $10^{-9}$ in very few iterations
and exhibits superior performance over the benchmarks. MM and $\mathsf{NLopt}$
are slower than SCQP by at least two orders of magnitude. Similar
to the first application, $\mathsf{DEoptim}$ and $\mathsf{GA}$ have
slow convergence or cannot obtain appropriate solutions within the
time limit. The results show the high efficiency of the proposed algorithm
in dealing with nonconvex objective functions.

Figure \ref{fig:kelly dimension} compares the CPU time of SCQP, MM,
and $\mathsf{NLopt}$ for different $N$. As expected, the proposed
SCQP algorithm shows a significant gain. When $N=50$, the convergence
speed achieved by SCQP is $\sim10.5$x higher than that of $\mathsf{NLopt}$.
This difference tends to grow in favor of SCQP as the problem size
increases. Besides, MM is more competitive than $\mathsf{NLopt}$
when more assets are considered. This may be because MM requires calling
the inner SOCP solver, and the setup time of the solver is less significant
compared with the solving time when the problem size is large. Table
\ref{tab:empirical order} also shows that SCQP enjoys the best scalability
with the lowest empirical time complexity order.

\subsection{Application III: Risk-constrained Markowitz Portfolio\label{subsec:Application-III}}

The third application we consider is the risk-constrained Markowitz
portfolio formulated as
\begin{equation}
\begin{aligned}\underset{\mathbf{w}\in\mathcal{W}}{\mathsf{minimize}}\,\,\, & \quad-\mathbf{w}^{\intercal}\boldsymbol{\mu}\\
\mathsf{subject\,\,to} & \quad\mathbf{w}^{\intercal}\boldsymbol{\Sigma}\mathbf{w}\le b.
\end{aligned}
\label{eq:risk-const Markowitz}
\end{equation}
Since its objective function satisfies the form of $\tilde{f}$, only
the inner loop of SCQP is required. Therefore, it is suitable for
evaluating the efficiency of the inner loop in dealing with mean-variance
constraints. Given that the equally weighted portfolio $\bar{\mathbf{w}}=\mathbf{1}/n$
achieves the risk $r=\bar{\mathbf{w}}^{\intercal}\boldsymbol{\Sigma}\bar{\mathbf{w}}$,
here we set the risk limit as $b=r$.

Problem (\ref{eq:risk-const Markowitz}) is in a standard SOCP form.
Therefore, we include SOCP solvers ($\mathsf{ECOS}$ and $\mathsf{MOSEK}$),
$\mathsf{NLopt}$, $\mathsf{DEoptim}$, and $\mathsf{GA}$ as benchmarks.
Figure \ref{fig:markowitz convergence} shows the convergence of different
methods in one realization of Problem (\ref{eq:risk-const Markowitz})
given $N=200$. The result indicates that SCQP is far better than
the benchmarks regarding convergence speed. Compared with Figure \ref{fig:robust convergence},
metaheuristic methods perform worse due to the difficulty in handling
nonlinear constraints.

Figure \ref{fig:markowitz dimension} compares the CPU time of all
methods except $\mathsf{DEoptim}$ and $\mathsf{GA}$ across different
problem sizes $N$. Compared with Figure \ref{fig:robust dimension},
we have two observations. First, SCQP is still much faster than the
benchmarks, especially when the dimension is high. This fact shows
that the inner loop handles the mean-variance constraints efficiently,
and taking advantage of sparsity is necessary. Second, SOCP solvers
and $\mathsf{NLopt}$ have consistent performance in optimizing different
SOCP problems. These observations are also corroborated by the similar
empirical time complexity order in Applications I and III, shown in
Table \ref{tab:empirical order}.

\subsection{Application IV: Long-term MSRP with Short-term Goals\label{subsec:Application-IV}}

\begin{figure*}
\begin{centering}
\hspace{1.5em}\subfloat[Convergence of algorithms in a realization with $N=200$. \label{fig:msrp convergence}]{\includegraphics[scale=0.3]{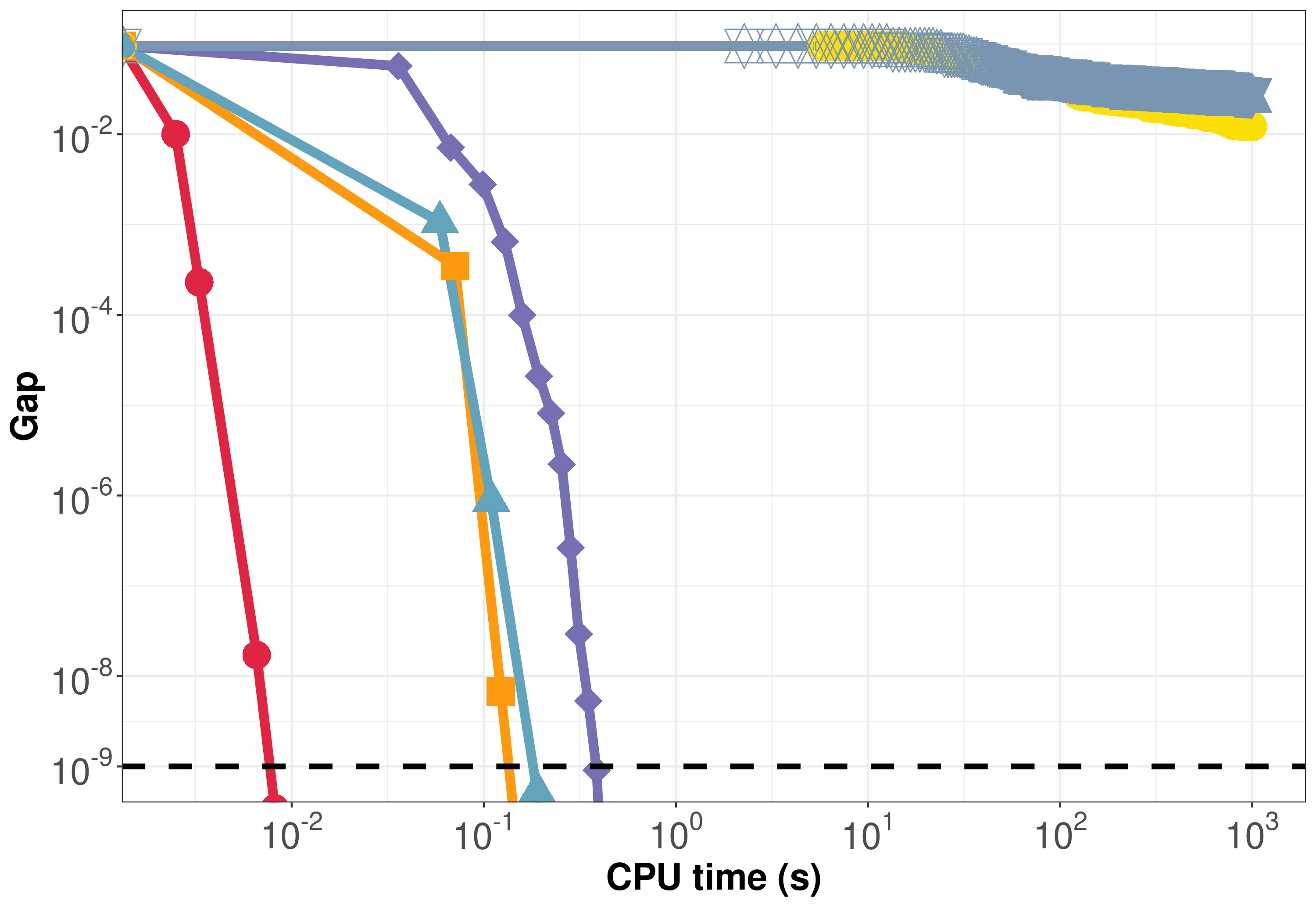}

}\hspace{1.7em}\subfloat[Timing results of algorithms with different $N$. \label{fig:msrp dimension}]{\includegraphics[scale=0.3]{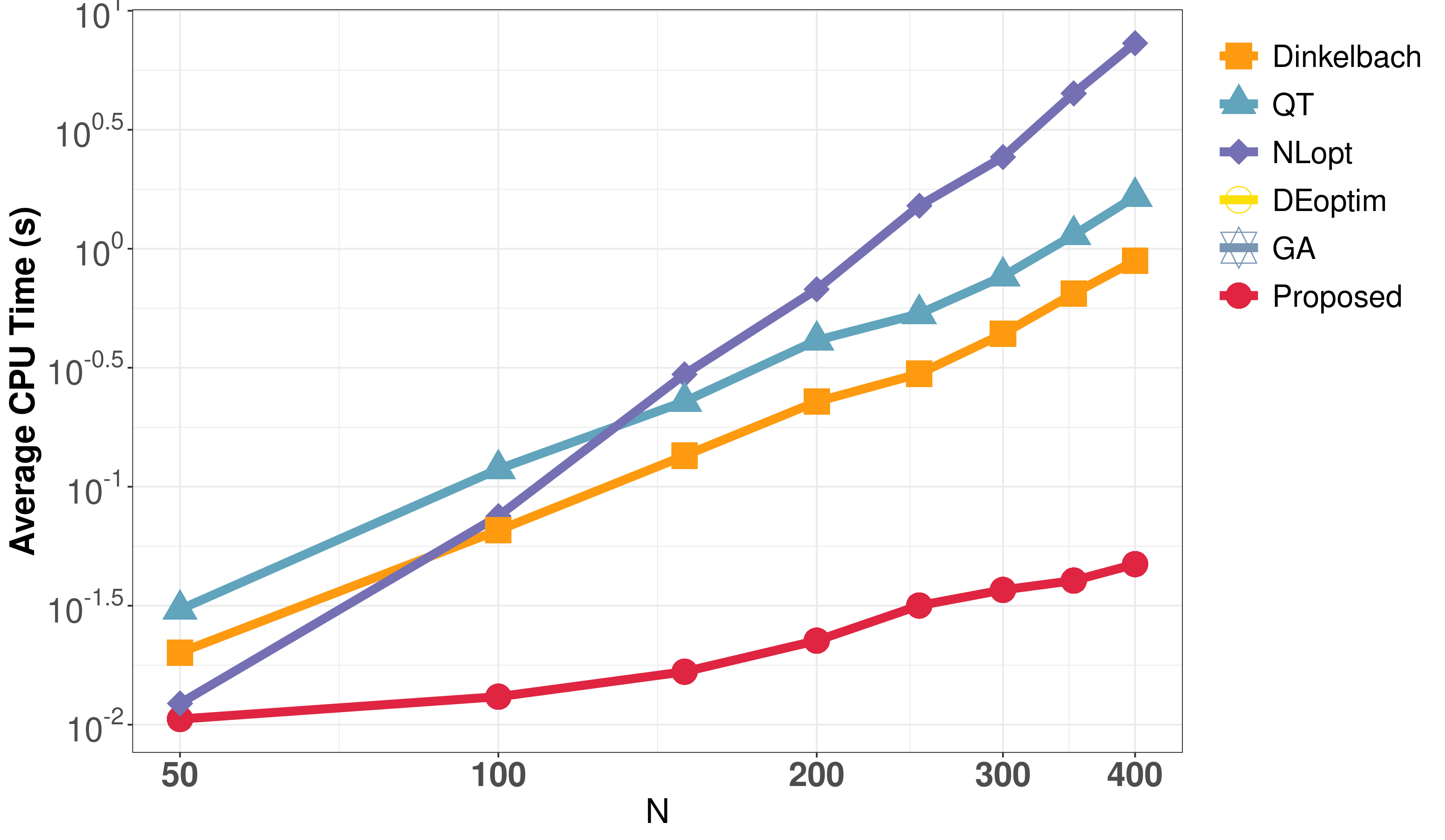}

}
\par\end{centering}
\caption{Numerical experiments on solving long-term MSRP with short-term goals
(\ref{eq:risk-constrained MSRP}).}
\end{figure*}
The fourth application we consider is the maximization of the long-term
Sharpe ratio along with the goals of running over the short-term market.
It is formulated as
\begin{equation}
\begin{aligned}\underset{\mathbf{w}\in\mathcal{W}}{\mathsf{minimize}}\,\,\, & \quad-\frac{\mathbf{w}^{\intercal}\boldsymbol{\mu}_{1}}{\sqrt{\mathbf{w}\mystrut^{\intercal}\boldsymbol{\Sigma}_{1}\mathbf{w}}}\\
\mathsf{subject\,\,to} & \quad\mathbf{w}^{\intercal}\boldsymbol{\mu}_{2}\ge a,\quad\mathbf{w}^{\intercal}\boldsymbol{\Sigma}_{2}\mathbf{w}\le b.
\end{aligned}
\label{eq:risk-constrained MSRP}
\end{equation}
Without loss of generality, $\boldsymbol{\mu}_{1}$ and $\boldsymbol{\Sigma}_{1}$
are estimated from the whole period of $5N$ trading days, while $\boldsymbol{\mu}_{2}$
and $\boldsymbol{\Sigma}_{2}$ are estimated from the latest $2N$
trading days. Given that the equally weighted portfolio $\bar{\mathbf{w}}=\mathbf{1}/n$
achieves the short-term expected return $e=\bar{\mathbf{w}}^{\intercal}\boldsymbol{\mu}_{2}$
and risk $r=\bar{\mathbf{w}}^{\intercal}\boldsymbol{\Sigma}_{2}\bar{\mathbf{w}}$,
we set $a=1.2e$ and $b=0.8r$, representing the interest in higher
return and lower risk than the short-term market.

Dinkelbach's algorithm and the quadratic transform (QT) solve Problem
(\ref{eq:risk-constrained MSRP}). Besides, $\mathsf{NLopt}$, $\mathsf{DEoptim}$,
and $\mathsf{GA}$ also serve as benchmarks. Figure \ref{fig:msrp convergence}
shows the convergence of different methods in one realization of Problem
(\ref{eq:risk-constrained MSRP}) given $N=200$. We observe that
the proposed SCQP reaches the gap of $10^{-9}$ faster than Dinkelbach's
algorithm and the quadratic transform by more than one order of magnitude.

Figure \ref{fig:msrp dimension} shows the average CPU time across
different problem sizes $N$. We observe that SCQP has the best performance,
consistently outperforming the benchmarks. Compared with the first
three applications that solely require the inner or outer loop, SCQP
takes longer solving time in this case because both loops are required.
Moreover, the performance of Dinkelbach's algorithm is fairly close
to that of the quadratic transform. One possible reason is that both
FP algorithms share the same algorithmic framework that recasts Problem
(\ref{eq:risk-constrained MSRP}) into a sequence of conic subproblems.
In addition, SCQP is far more efficient than $\mathsf{NLopt}$ when
the dimension increases. The empirical time complexity order in Table
\ref{tab:empirical order} further supports our findings.

\section{Conclusion\label{sec:conclusion}}

In this paper, we proposed and analyzed a successive QP algorithm
for general MVP optimization. The main advantage is that no matter
what problem structures are contained in different MVPs, the proposed
algorithm only requires solving a sequence of QP surrogate problems
which already have well-developed efficient solvers. In addition,
by exploiting the underlying sparsity pattern of this algorithm, we
proposed the fast implementation that can further reduce the computational
cost. The theoretical convergence analysis has been established, and
comprehensive experiments reveal that the proposed algorithm has a
higher convergence speed and better scalability than the state-of-the-art
methods.

\appendix{}

\subsection{Proof for Proposition \ref{prop:active set strategy} \label{apx: proof active set}}

We begin with the following technical lemma.
\begin{lem}
\label{lem:Decreasing_size}For any two adjacent iterations in Algorithm
\ref{alg:Active-set-strategy}, we have $(\bar{L}^{k+1}\cup\bar{U}^{k+1})\subset(\bar{L}^{k}\cup\bar{U}^{k})$
together with $q(\mathbf{w}^{k+1})<q(\mathbf{w}^{k})$.
\end{lem}
\begin{IEEEproof}
$\mathbf{w}^{k}$ is the solution of Problem (\ref{eq:ASM_sub}) with
the KKT condition
\begin{equation}
\beta_{i}^{l}\geq0,\:\:\beta_{i}^{u}\geq0,\:i\notin\bar{L}^{k}\cup\bar{U}^{k},
\end{equation}
so the ``violated'' assets with $\beta_{i}^{l}<0$ or $\beta_{i}^{u}<0$
must belong to the working set. Relaxing the strict equality bound
constraints on these variables results in a smaller working set, i.e.,
$(\bar{L}^{k+1}\cup\bar{U}^{k+1})\subset(\bar{L}^{k}\cup\bar{U}^{k})$.
Thus, we have
\begin{equation}
\begin{aligned} & \left\{ \mathbf{w}\left|\begin{array}{c}
\mathbf{A}\mathbf{w}=\mathbf{b},\quad l_{i}\le w_{i}\le u_{i},\:i\notin\bar{L}^{k}\cup\bar{U}^{k}\\
w_{i}=l_{i},\:i\in\bar{L}^{k},\quad w_{i}=u_{i},\:i\in\bar{U}^{k}
\end{array}\right.\right\} \\
\subset & \left\{ \mathbf{w}\left|\begin{array}{c}
\mathbf{A}\mathbf{w}=\mathbf{b},\quad l_{i}\le w_{i}\le u_{i},\:i\notin\bar{L}^{k+1}\cup\bar{U}^{k+1}\\
w_{i}=l_{i},\:i\in\bar{L}^{k+1},\quad w_{i}=u_{i},\:i\in\bar{U}^{k+1}
\end{array}\right.\right\} ,
\end{aligned}
\end{equation}
which means $\mathbf{w}^{k}$ is also a feasible point of the subproblem
(\ref{eq:ASM_sub}) in the $(k+1)$th iteration. Also, $\mathbf{w}^{k}\neq\mathbf{w}^{k+1}$,
or we would not have ``violated'' assets. Therefore, each iteration
must decrease the objective function, i.e., $q(\mathbf{w}^{k+1})<q(\mathbf{w}^{k})$.
\end{IEEEproof}
We are now ready to prove Proposition \ref{prop:active set strategy}.
First of all, the proposed algorithm is convergent. This is guaranteed
by the monotone decrease of the size of the working set, according
to Lemma \ref{lem:Decreasing_size}. As the minimum size of the working
set is bounded by zero (i.e., optimality achieved), the algorithm
converges to the optimal solution within finite steps. At each iteration,
at least one index is dropped from the working set. Hence, the number
of iterations is bounded by $|\bar{L}^{0}\cup\bar{U}^{0}|$.

\subsection{MM Algorithm for Kelly Portfolio \label{apx: MM for Kelly}}

To solve Problem (\ref{eq:Kelly}), we first apply the quadratic transform
on the ratio term and reformulates it as
\begin{align}
\underset{\mathbf{w}\in\mathcal{W}}{\mathsf{minimize}}\;\; & (-\log\left(1+\mathbf{w}^{\intercal}\boldsymbol{\mu}\right)+a_{1}\sqrt{\mathbf{w}\mystrut^{\intercal}\boldsymbol{\Sigma}\mathbf{w}}\nonumber \\
 & -\frac{a_{1}^{2}}{2}\left(1+\mathbf{w}^{\intercal}\boldsymbol{\mu}\right)^{2})\label{eq:Kelly-QT}
\end{align}
where $a_{1}$ is an auxiliary variable iteratively updated by
\begin{equation}
a_{1}=\sqrt{\left(\mathbf{w}^{k}\right)^{\intercal}\boldsymbol{\Sigma}\mathbf{w}^{k}}/\left(1+\boldsymbol{\mu}^{\intercal}\mathbf{w}^{k}\right)^{2}.
\end{equation}
As shown by \cite{shen2019optimization}, (\ref{eq:Kelly-QT}) is
an upper-bound problem of the primal problem (\ref{eq:Kelly}). Further,
we apply the first-order Taylor expansion to the nonconvex quadratic
term of the objective function in (\ref{eq:Kelly-QT}). It constructs
a convex upper-bound problem of (\ref{eq:Kelly-QT}) given by
\begin{align}
\underset{\mathbf{w}\in\mathcal{W}}{\mathsf{minimize}}\;\; & (-\log\left(1+\mathbf{w}^{\intercal}\boldsymbol{\mu}\right)+a_{1}\sqrt{\mathbf{w}\mystrut^{\intercal}\boldsymbol{\Sigma}\mathbf{w}}\nonumber \\
 & -a_{1}^{2}\left(\boldsymbol{\mu}+\mathbf{a}_{2}\right)^{\intercal}\mathbf{w}+c)\label{eq:Kelly QT+MM}
\end{align}
where $c$ is a constant, and $\mathbf{a}_{2}$ is another auxiliary
variable computed as
\begin{equation}
\mathbf{a}_{2}=\boldsymbol{\mu}\boldsymbol{\mu}^{\intercal}\mathbf{w}^{k}.
\end{equation}
Over all, this algorithm solves Problem (\ref{eq:Kelly}) by solving
a sequence of upper-bound problems (\ref{eq:Kelly QT+MM}), and thus
it can be interpreted as an MM algorithm.

\bibliographystyle{IEEEtran}
\bibliography{meanVarPortfolio_arxiv}

\end{document}